\documentclass[twocolumn]{aa}
\usepackage{graphicx}
\usepackage{txfonts}
\newcommand {\apgt} {\ {\raise-.5ex\hbox{$\buildrel>\over\sim$}}\ }
\newcommand {\aplt} {\ {\raise-.5ex\hbox{$\buildrel<\over\sim$}}\ }

\begin{document}
   \title{Testing the inverse-Compton catastrophe scenario
          in the intra-day variable blazar \object{S5\,0716+71}}

   \subtitle{II. A search for intra-day variability at
             millimetre wavelengths with the IRAM 30\,m telescope}

   \titlerunning{A search for intra-day variability at
             millimetre wavelengths with the IRAM 30\,m telescope}

   \author{I. Agudo
          \inst{1}
          \and
          T.P. Krichbaum
	  \inst{1}
          \and
          H. Ungerechts
	  \inst{2}
          \and
          A. Kraus
	  \inst{1}
          \and
          A. Witzel
	  \inst{1}
          \and
           E. Angelakis
	  \inst{1}
          \and
           L. Fuhrmann
	  \inst{1,3,4}
          \and
           U. Bach
	  \inst{1,4}
          \and
           S. Britzen
	  \inst{1}
          \and
           J. A. Zensus
	  \inst{1}
          \and
           S. J. Wagner
	  \inst{5}
          \and
           L. Ostorero
	  \inst{5,6}
          \and
           E. Ferrero
	  \inst{5}
          \and
           J. Gracia
	  \inst{7}
          \and
           M. Grewing
	  \inst{8}
          }

   \offprints{I. Agudo, \email{iagudo@mpifr-bonn.mpg.de}}

   \institute{Max-Planck-Institut f\"ur Radioastronomie,
              Auf dem H\"ugel, 69,
              D-53121, Bonn, Germany
         \and
             Instituto de Radio Astronom\'{\i}a Milim\'etrica,
	     Avenida Divina Pastora, 7, Local 20,
             E-18012, Granada, Spain
         \and
             Dipartimento di Fisica, Universit\`a di Perugia,
	     via A. Pascoli, 06123 Perugia, Italy
         \and
             INAF, Osservatorio Astronomico di Torino,
	     via Osservatorio 20, 10025 Pino Torinese (TO), Italy
         \and
             Landessternwarte Heidelberg-K\"onigstuhl, K\"onigstuhl,
             D-69117, Heidelberg, Germany
         \and
	     Tuorla Observatory, University of Turku, V\"ais\"al\"antie
	     20, 21500 Piikki\"o, Finland
         \and
	     Department of Physics, University of Athens,
	     Panepistimiopolis, 157 84 Zografos,
	     Athens, Greece
         \and
             Institut de Radio Astronomie Milim\'etrique,
	     300 Rue de la Piscine, Domaine Universitaire
	     de Grenoble, St. Martin d'H\`eres, F-38406, France\\
             }

   \date{Received December 6, 2005; accepted May 30, 2006}

   \abstract{We report on a densely time sampled polarimetric flux density
   monitoring of the BL~Lac object  \object{S5\,0716+71} at 86\,GHz
   and 229\,GHz.
   The source was observed with the IRAM 30\,m telescope at Pico
   Veleta within a coordinated multi-frequency observing campaign,
   which was centred around a 500\,ks INTEGRAL observation during November
   10 to 16, 2003.  The aim of this campaign was to search
   for signatures of inverse-Compton catastrophes through the
    observation of the broad-band variability of the source.
   At 86\,GHz,  S5\,0716+71 showed no intra-day variability,
   but showed remarkable inter-day variability with a flux density
   increase of $34$\,\% during the first four observing days,
    which can not be explained by source extrinsic causes.
   At this frequency,
   making use of a new calibration strategy, we reach a relative
   rms accuracy of the flux density measurements of $1.2$\,\%.
   Although the flux density variability at 229\,GHz was consistent
   with that at 86\,GHz, the larger measurement errors at
   229\,GHz do not allow us to detect, with high confidence,
    inter-day variations at this frequency. At 86\,GHz, the
   linear polarization fraction of  S5\,0716+71 was unusually large
    $\rm{(}15.0\pm1.8\rm{)}\,\%$.
   Inter-day variability in linear polarization at 86\,GHz,
   with significance level $\apgt  95\,\%$;
   $\sigma_{P}/<P>=15$\,\% and $\sigma_{\chi}=6^{\circ}$, was observed
   during the first four observing days.
   From the  total flux density variations at the synchrotron turnover
   frequency ($\sim 86\,\rm{GHz}$) we compute an apparent brightness
   temperature $T_{\rm{B}}^{\rm{app}} > 1.4 \times {10}^{14}$\,K
    at a redshift of 0.3, which exceeds  by two orders of magnitude
   the inverse-Compton limit.
   A relativistic correction for $T_{\rm{B}}^{\rm{ app}}$ with a
   Doppler factor $\delta > 7.8$ brings  the observed brightness
   temperature down to the inverse Compton limit.
   A more accurate lower limit of $\delta > \rm{ 14.0}$, consistent with
   previous  estimates from VLBI observations, is obtained
   from the comparison of  the 86\,GHz synchrotron  flux density
   and  the upper limits for the synchrotron self-Compton
    flux density obtained from the INTEGRAL observations.
   The relativistic beaming of the emission by this high Doppler
   factor explains the non-detection of ``catastrophic"
   inverse-Compton avalanches by INTEGRAL.

   \keywords{galaxies: active --
             galaxies: BL~Lacertae objects: general --
             galaxies: BL~Lacertae objects: individual:  S5\,0716+71 --
	     radio continuum: galaxies --
	     radiation mechanisms: non-thermal --
	     polarization
               }
   }

   \maketitle
%

\section{Introduction}
\label{int}

The effect of intra-day variability (IDV) of radio loud active
galactic nuclei (AGN) in the radio bands was discovered in 1986
(Witzel et al.~\cite{Wit86}; Heeschen et al.~\cite{Hee87}) and
since then led to controversial discussions about its physical
origin.
Most of the objects presenting radio IDV appear
compact -when are observed in the radio and optical bands-,
their relativistic jets are highly core-dominated on VLBI
scales and exhibit high brightness temperatures (e.g.
Wagner \& Witzel~\cite{Wag95}).
IDV is a common phenomenon in extragalactic flat
spectrum radio sources and is observed, at centimetre wavelengths,
in about 10\% to 25\% of all objects of this class
(Quirrenbach et al.~\cite{Qui92};
Kedziora-Chudczer et al.~\cite{Ked01}; Lovell et al.~\cite{Lov03}).
The ``classical IDV" (of type II, as defined in Heeschen et al.
\cite{Hee87}) is characterised by variability amplitudes
$\aplt 20$\,\% and variability time-scales between 0.5\,days
and 2\,days.
In parallel to the variability of the total flux density,
similar or even faster variations of the linear polarization
(Quirrenbach et al.~\cite{Qui89}; Kraus et al.
\cite{Kra99a},~\cite{Kra99b},~\cite{Kra03}) and of the
circular polarization (Macquart et al.~\cite{Mac00})
are observed.
Intensity and polarization variations can be correlated or
anti-correlated (e.g. Wagner et al.~\cite{Wag96};
 Macquart et al.~\cite{Mac00}; Qian et al.~\cite{Qia02}).

Observations of IDV sources in the radio and millimetre
bands are particularly important, since they reveal the
highest apparent brightness temperatures
($T_{\rm{B}} \propto \nu^{-2}$; e.g. Readhead~\cite{Rea94}).
However, the observed radio emission might also be severely
affected by interstellar scintillation (ISS), which is unavoidable
because of the small intrinsic sizes of IDV sources (Rickett
\cite{Ric90}).
It is known that, for high Galactic latitude blazars,
the amplitude of IDV due to the
effect of interstellar scintillation rapidly decreases
with frequency for $\nu \apgt 5$\,GHz (e.g. Rickett~\cite{Ric90};
Rickett et al.~\cite{Ric95}; Beckert et al.~\cite{Bec02}).
In such case and for a source whose size is constant with
frequency, the ISS-induced variability amplitude scales as
$\propto \nu^{-2}$ (e.g. Beckert et al.~\cite{Bec02}).
Hence, any rapid and large amplitude source
variability seen at millimetre (or shorter) wavelengths should
not be strongly affected by ISS.
Therefore, monitoring of IDV sources at mm-bands is particularly
important to determine whether the observed IDV is
mainly source extrinsic (scintillation), due to source
intrinsic processes, or a mixture of both.

\object{S5\,0716+71} (hereafter \object{0716+714})
is an extremely active BL~Lac
object which varies on time scales from less than one hour to
months, from radio to X-rays (e.g. Wagner et al.~\cite{Wag96},
Kraus et al.~\cite{Kra03}, Raiteri et al.~\cite{Rai03}).
Although the optical spectrum of \object{0716+714} appears
featureless, even with 4\,m class telescopes, the
absence of any signature of a host galaxy (in deep images) sets
a lower limit to its redshift of $z>0.3$ (Wagner et al.~\cite{Wag96}).
Therefore, only lower limits to the brightness temperature
of the source can be derived. The strongest constraint
on the brightness temperature of this source comes from the
observed short variability time-scales at radio wavelengths,
which sets, via the causality argument, an upper limit
to the source size of a few tens of micro-arcseconds.
During the last decade the source has been observed repeatedly in
simultaneous centimetre wavelength campaigns (Wagner et al.
\cite{Wag90},~\cite{Wag96}; Otterbein et al.~\cite{Ott99};
Kraus et al.~\cite{Kra03}; Raiteri et al.~\cite{Rai03}),
during which it typically showed IDV.
\object{0716+714} is the only IDV source which has shown
simultaneous variability-time-scale transitions in the
radio and optical bands, which have been interpreted as evidence
for source intrinsic variability (Quirrenbach et al.~\cite{Qui91};
Wagner \& Witzel~\cite{Wag95}; Qian et al.~\cite{Qia96}).
The correlation between the radio spectral index
and the optical flux suggested that the radiation was
produced by the same particle population (Qian et al.~\cite{Qia96}).
Additional arguments for the relevance of the intrinsic origin of
the IDV in \object{0716+714} come from the variability
amplitude, which increases from radio to optical, and the recently
detected IDV at a wavelength of 9\,mm
(Kraus et al.~\cite{Kra03}).
Both effects contradict expectation for the frequency dependence
of ISS.

If IDV is produced by the intrinsic properties of AGN,
very small angular sizes of the emitting regions
(of a few micro-arcseconds) are inferred from their short
variability time-scales.
This usually implies apparent brightness temperatures several
orders of magnitude larger than the inverse-Compton (IC)
limit ($\aplt 10^{12}$\,K, Kellermann \& Pauliny-Toth~\cite{Kel69};
Readhead~\cite{Rea94}; Kellermann~\cite{Kel03}).
For incoherent synchrotron
sources of radiation, this limit could be violated only during
short time ranges, because the high energy photon densities
at the IC-limit should lead to the rapid cooling of the
emitting region through the IC scattering of the synchrotron
radiation. This process is known as the \emph{inverse-Compton
catastrophe}.
The relativistic beaming of the radiation coming from the
emitting region (Rees~\cite{Ree66}) has been proposed as a
likely cause to explain the systematic violation of the
IC-limit (e.g. Wagner \& Witzel~\cite{Wag95}).
Other scenarios, as coherent synchrotron radiation, which
is an unconventional but not impossible process in AGN, have
been proposed (e.g. Benford~\cite{Ben92}).
However, the inverse-Compton catastrophe scenario can not
still be ruled out (Wagner \& Witzel~\cite{Wag95}).

A coordinated broad-band observing campaign
-centred around a 500\,ksec INTEGRAL\footnote{INTErnational
Gamma-Ray Astrophysics Laboratory} observation
during November 10 to 16, 2003- was performed to search
for signatures of inverse-Compton catastrophes on a flaring
state of \object{0716+714} through the correlation of the broad-band
variability of the source.
The ground-based observations were performed
with the VLBA, the Effelsberg 100\,m, the Mets\"ahovi 13.7\,m,
the IRAM 30\,m, the JCMT, the HHT, and the Kitt Peak 12\,m
telescopes and the optical-IR telescopes of the
WEBT\footnote{Whole Earth Blazar Telescope} collaboration.
The first analysis of the broad-band (from radio to soft
$\gamma$-rays) observations (Ostorero et al.~\cite{Ost06})
do not show obvious correlation between the intra-day optical
variability and the inter-day radio variability displayed by
the source. Although the apparent brightness temperatures of
\object{0716+714} largely exceeded the IC-limit
at radio wavelengths, no evidence of IC avalanches
was seen in the INTEGRAL data.

In this paper we report on the results and implications
of the first polarimetric millimetre wavelength IDV observations
of \object{0716+714}, which were performed (within the above-mentioned
campaign) with the IRAM 30\,m radio telescope during the period
November 10 to 18, 2003.
A combined discussion of the radio, millimetre and sub-millimetre
single dish data set will be presented by Fuhrmann et al.~(\cite{Fuh06}).
Forthcoming papers will describe more detailed analysis of
the optical observations and more sophisticated theoretical
modelling of the broad-band data set.


\section{Observations and data reduction}
\label{obsred}

\subsection{Observations}
\label{obs}

The observations presented here were specially planned to achieve
accurate measurements of the total flux density
with a typical temporal resolution better than 1\,h.
To achieve this goal, we observed \object{0716+714} and a number of suitable
calibrators continuously for $\sim 18$\,h on each day during November
10 to 18, 2003, using the IRAM 30\,m telescope at Pico Veleta (Granada,
Spain).
The observations were interrupted by intervals where the elevation of
the program source was lower than $20^{\circ}$ or when the weather
conditions were inappropriate for 86\,GHz observations. Under these
conditions, accurate antenna and gain calibrations were not possible.

The receiver-cabin optic system of the IRAM 30\,m telescope is
optimised to observe with up to four different receivers at the same time
(Wild~\cite{Wil99}).
We made use of the A100, A230, B100 and B230 heterodyne receivers
of the telescope, which simultaneously observed at both
86\,GHz (A100 and B100, at $\lambda \simeq 3.5$\,mm, where $\lambda$ is
the wavelength) and 229\,GHz (A230 and B230, at $\lambda \simeq 1.3$\,mm).
This standard observing set-up uses a grid to divide the incoming
signal into two orthogonal linear polarizations and then two
Martin-Pupplet interferometers (one per polarization) to split the
86\,GHz and the 229\,GHz frequency bands.
Hence, at each observing band, the A- and B-receivers were sensitive
to two orthogonal linear polarizations.
This allowed us to obtain linear polarization information at 86\,GHz.
Receiver A230 was not stable enough to provide reliable measurements
and hence we lost the polarization information at 229\,GHz.
The observing bandwidths at 86\,GHz and 229\,GHz were 0.5\,GHz and
1\,GHz, and the typical single side band system temperatures
were $\sim 100$\,K and $\sim 400$\,K, respectively.

All the measurements were performed in ``beam-switching" mode,
using a chopper wheel to subtract the emission contribution from
the sky.
During the continuous chopping (at a frequency of $\approx7$\,Hz)
the telescope beam was moved to a 70\hbox{$^{\prime\prime}$} off-position.
The individual measurements were performed by repeated cross-scans
over the source position in azimuth and elevation.
The number of sub-scans in each measurement and for each source was
chosen according to the flux density of the source, with more
sub-scans on the fainter objects.
For technical reasons, the number of sub-scans were multiples of
four and ranged between 4 and 16.
\object{0716+714} was observed with a ``duty cycle" of about
1 to 2 measurements every 60\,minutes.
Also six bright and point-like extragalactic
radio sources (\object{S5\,0212+735}, \object{S5\,0633+734},
\object{S5\,0836+710}, \object{S4\,1642+690}, \object{S5\,1803+784}, \object{S5\,1928+738}, which will be called hereafter by their
catalogue codes) were observed with a duty cycle
of at least one measurement every hour.
Their known variability characteristics at
cm-wavelengths and location on the sky allowed us to assume
that they were non-variable on time-scales of days and
so they were suitable as secondary calibrators\footnote{In \S \ref{sres}
we show that this assumption was true for all the observed
extragalactic calibrators but \object{1642+690}}.
For the determination of the absolute flux density scale,
we observed two planets (\object{Mars} and \object{Uranus})
at least once per day and we also included two bright and
compact H\,II regions (\object{W3\,OH}, \object{K3-50A}) and the
planetary nebula \object{NGC\,7027} in the measurement
cycle every 4\,h to 5\,h.

The frequent observation of the target source and the
primary and secondary calibrators, in combination with
our non-standard data reduction procedure, enabled us to
calibrate the measurements with unprecedented precision.
This was possible thanks to the accurate determination
of the dependence of calibration on elevation and time
from the measurements of the calibrators.
This method has been successfully proven on previous
observations of IDV (e.g. Quirrenbach et al.\cite{Qui92};
Kraus et al.~\cite{Kra03}), which achieved relative calibration
accuracies better than 1\,\% at radio frequencies.
The method also allows for the empirical determination of
the internal calibration errors without the need of a
detailed knowledge of the individual sources of error
affecting the measurements.


\subsection{Total flux density data reduction}
\label{sred}

The initial step of the data calibration used a method equivalent
to the ``chopper-wheel\footnote{Note that the ``chopper-wheel"
calibration procedure (which at the 30\,m telescope does not
use a chopper wheel) is totally unrelated to the
``beam-switching" mode (which uses the chopper wheel).
This paradoxical nomenclature has been maintained
for historical reasons.}" calibration
(e.g. Kutner \& Ulich~\cite{Kut81}) to translate the
detected power (measured in arbitrary units) into calibrated
and opacity corrected antenna temperatures ($T_{A}^{*}$ in K).
The ``chopper-wheel" procedure uses measurements of a hot and a
cold load, both at known temperatures, and the sky temperature
to compute the calibration factor for each measurement.
At the IRAM 30\,m telescope, this calibration factor
takes into account the total absorption and thermal
emission of the atmosphere, which are computed by an
atmospheric radiative transfer model (see Kramer~\cite{Kra97}
and references therein for details).
For our adopted observing set-up at the IRAM 30\,m telescope,
this is the standard initial calibration, which is performed
in ``real time" by the software of the telescope.

The post-observation data reduction was
performed following an incremental strategy in different
stages, further improving the calibration accuracy after
each of them.
After each correction the data were
inspected and edited following different criteria,
depending on the previous correction.
The rms of the normalised measurements of the secondary
calibrators are listed in Table~\ref{rmscal} to show
how much each of the corrections improved the calibration.
Here we explain the corrections applied on each of the
data reduction stages:

Stage 1:
We first averaged for each scan independently the sub-scans
in azimuth and in elevation.
Then, we fitted a Gaussian profile to the average of each of
the two scanning directions.
The amplitude of the Gaussian corresponds to $T_{A}^{*}$
in both directions.

Stage 2:
Using the measured FWHM of the telescope beam
($\approx 28$\hbox{$^{\prime\prime}$} at 86\,GHz and
$\approx 11$\hbox{$^{\prime\prime}$} at 229\,GHz), we were able
to correct the amplitudes measured on the two directions
for the observed (small, typically $\aplt 3$\hbox{$^{\prime\prime}$})
telescope pointing errors.
These corrections were typically smaller than 5\,\% at
86\,GHz and 25\,\% at 229\,GHz.
After this correction, we averaged the measurements from the
azimuth and elevation directions to produce a single
$T_{A}^{*}$ measurement per receiver and scan.

Stage 3:
Subsequently, we corrected for elevation dependent effects
on the data, i.e. caused by gravitational deformation of the
telescope dish.
For this purpose, we used the antenna gain curves given
by Greve et al.~(\cite{Gre98}). After this correction
($\aplt 2\,\%$ at 86\,GHz and $\aplt 5\,\%$ at 229\,GHz),
the data did not show any significant elevation dependence.

Stage 4:
After that, we removed the most
obvious systematic time-dependent variations,
i.e. gain decreases (which were $\aplt 10\,\%$ at
86\,GHz and $\aplt 30\,\%$ at 229\,GHz) mainly
produced by abrupt ambient temperature changes.
The corresponding gain corrections were computed by making
use of the densely time-sampled measurements of all our secondary
calibrators besides \object{1642+690}, and assuming that they
were constant during our observations. The latter was
confirmed by our variability analysis (see \S~\ref{sres}).

Stage 5:
At 86\,GHz, it was also possible to correct for a small
relative gain difference of receivers A100 and B100
($\simeq$\,1.035\,\%), detected by comparing good quality
measurements of unpolarized calibrators.
After that, the averaging of the results from A100 and
B100 enabled to reduce the uncertainties on the
determination of the $T_{A}^{*}$ measurements at 86\,GHz.
An accurate relative calibration of the A100 and B100
gains is also required to measure the linear polarization
properties of the polarized sources at 86\,GHz (see
Appendix~\ref{appA} and \S~\ref{pred}).
Due to the lack of reliable measurements from the A230
receiver, these data improvements could not be performed
for the 229\,GHz data.

Stage 6:
To further improve our calibration, an additional
time-dependent correction, similar to the
one explained on stage 4, was applied to correct
for residual systematic time variations of the overall
gain.
The corrected gain variations were $\aplt 5\,\%$ at
86\,GHz and $\aplt 20\,\%$ at 229\,GHz on this stage.

Stage 7:
Then we derived the Kelvin-to-Janksy
conversion factors ($C_{\rm{KJ}}$) by comparing
the measurements of \object{Mars}, \object{Uranus}, \object{W3\,OH}, \object{K3-50A}
and \object{NGC\,7027} (which are all of them unpolarized)
with their assumed absolute surface
brightnesses (Table~\ref{primcal}).
The final values,
$C_{\rm{KJ}}(\rm{86\,GHz})= (6.37\pm0.04)$\,Jy/K and
$C_{\rm{KJ}}(\rm{229\,GHz})= (9.3\pm0.3)$\,Jy/K are the result
of the average of the different Kelvin-to-Janksy conversion
factors computed from each of the above calibrators.
The errors in $C_{\rm{KJ}}$ take into account the uncertainties
of our measurements of the primary calibrators,
but not the possible inaccuracies on the
assumed values of their surface brightnesses.
$C_{\rm{KJ}}(\rm{86\,GHz})$ and $C_{\rm{KJ}}(\rm{229\,GHz})$
were finally applied to all the data to obtain absolute flux
density measurements ($S$ in Jy) for \object{0716+714}
and the secondary calibrators.

Note that the absolute surface brightnesses of
\object{Uranus}, \object{W3\,OH}, \object{K3-50A} and
\object{NGC\,7027} were calibrated
relative to those of \object{Mars} (Table~\ref{primcal};
Griffin \& Orton~\cite{Gri93}; Kramer~\cite{Kra97}).
Hence, an additional 5\,\% absolute calibration error,
coming from uncertainties in the true martian
temperatures (Griffin \& Orton~\cite{Gri93}), would
have to be quadratically added to our absolute flux
density results.
However, as this does not affect our main relative
variability analyses -based on measurements
scaled to the same Kelvin-to-Janksy factor- the
extra 5\,\% error has not been added, unless
explicitly indicated for each particular
absolute flux density measurement.

\begin{table}
\caption[]{{rms of the normalised measurements of the secondary
           calibrators (but \object{1642+690}) after each of the
	   data reduction stages.}}
\begin{flushleft}
\footnotesize
\begin{tabular} {lcccccccc}
\hline\noalign{\smallskip}
         &  \multicolumn{8}{c}{Data reduction stage}\\
\hline\noalign{\smallskip}
 Frequency  & 1 & 2 & 3 & 4 & 5 & 6 & 7 & 8 \\
            & \multicolumn{8}{c}{(\%)} \\
\hline\noalign{\smallskip}
86\,GHz  & 12 & 11 & 5 & 4 & 4 & 1.4 & 1.2 & 1.2 \\
229\,GHz & 120 &  100 & 30 & 19 & ... & 16 & 16 & 16 \\
\noalign{\smallskip}
\hline
\end{tabular}
\end{flushleft}
\label{rmscal}
\end{table}

\begin{table*}
\caption[]{Absolute surface brightnesses (in Jy/beam) of
           \object{Mars}, \object{Uranus}, \object{W3\,OH},
	   \object{K3-50A} and \object{NGC\,7027} adopted
           for the computation of the { Kelvin-to-Janksy}
	   conversion factors { of the 30\,m telescope}
	   at 86\,GHz and 229\,GHz.}
\begin{flushleft}
\footnotesize
\begin{tabular} {lcccccc}
\hline\noalign{\smallskip}
         &  \multicolumn{6}{c}{U.~T. days of Nov. 2003}\\
\hline\noalign{\smallskip}
 Source  & 11 & 12 & 13 & 14 & 15 & 16 \\
\hline\noalign{\smallskip}
\multicolumn{7}{c}{$86$\,GHz}\\
\hline\noalign{\smallskip}
\object{Mars}$^{\rm{a}}$      & 149.75 & 146.88 & 144.08 & 141.33 & 138.65 & 136.02 \\
\object{Uranus}$^{\rm{a}}$    &   7.06 &   7.05 &   7.04 &   7.03 &   7.01 &   7.01 \\
\object{W3\,OH}$^{\rm{b}}$    & \multicolumn{6}{c}{3.94} \\
\object{K3-50A}$^{\rm{b}}$   & \multicolumn{6}{c}{6.29} \\
\object{NGC\,7027}$^{\rm{b}}$ & \multicolumn{6}{c}{4.71} \\
\hline\noalign{\smallskip}
\multicolumn{7}{c}{$229$\,GHz}\\
\hline\noalign{\smallskip}
\object{Mars}$^{\rm{a}}$      & 691.16 & 683.33 & 675.51 & 666.71 & 659.92 & 652.17 \\
\object{Uranus}$^{\rm{a}}$    &  32.88 &  32.83 &  32.77 &  32.72 &  32.67 &  32.61 \\
\object{W3\,OH}$^{\rm{b}}$    & \multicolumn{6}{c}{6.48} \\
\object{K3-50A}$^{\rm{b}}$   & \multicolumn{6}{c}{6.98} \\
\object{NGC\,7027}$^{\rm{b}}$ & \multicolumn{6}{c}{3.68} \\

\noalign{\smallskip}
\hline
\end{tabular}
\end{flushleft}
$^{\rm{a}}$Computed from the expressions given in
           Kramer~(\cite{Kra97}), and references therein,
	   { which take into account the angular size of
	   the planets. The values given for \object{Uranus} were
	   computed from an empirical model which used
	   \object{Mars} as primary calibrator
	   (Griffin \& Orton~\cite{Gri93}).}

$^{\rm{b}}$Obtained from averages of measurements performed
           during several years - before 1996 - at the
	   IRAM 30\,m telescope. These measurements were
	   themselves calibrated through observations of planets.
	   { The sizes of these calibrators at 1.2\,mm are given
	   in Lisenfeld et al.~(\cite{Lis00}).}
\label{primcal}
\end{table*}

{ Stage 8:}
{ Finally, we performed an empirical estimate of
the internal calibration errors, including those that could
not be taken into account previously (mainly those affecting
the gain corrections of data reduction steps 4, 5 and 6).
This {\it -a posteriori-} error estimate was performed by
computing the} rms of the { normalized} flux density
measurements { of the assumed non-variable} secondary
calibrators { (all of them except} \object{1642+690}),
{  which resulted} $\rm{rms}_{86}=1.2$\,\% and
$\rm{rms}_{229}=16$\,\% for the 86\,GHz and 229\,GHz data,
respectively.
{ These quantities were finally added in quadrature to
the error of each individual measurement on stage 7, which were
initially computed from the 1-$\sigma$ uncertainty estimates
of the amplitude of the Gaussian profiles and were then
propagated through the different averages and corrections
of the data reduction procedure.}
The final { relative} errors of each individual measurement
{ were} typically of $\sim 2$\,\% at 86\,GHz and
$\sim 18$\,\% at 229\,GHz for \object{0716+714}.
{ As indicated above, an additional 5\,\% factor should
still be added quadratically to account for the absolute
flux density errors.}

{ Note that the small receiver-temperature
changes monitored during the observations did not allow flux
density fluctuations larger than $\sim 0.5$\,\% either at
86\,GHz or at 229\,GHz.
Hence, the residual $\rm{rms}_{86}$ and $\rm{rms}_{229}$
statistical uncertainties are most likely resulting from
the small and rapid atmospheric fluctuations between
independent measurements, which due to the limited time
sampling could not be corrected.}

{ It is worth to stress} that the small $\rm{rms}_{86}=1.2$\,\%
characterises { the excellent performance of the IRAM 30\,m
telescope and the stability of 86\,GHz observations at the
telescope site.}
This also demonstrates the ability for future high accuracy
IDV studies at millimetre wavelengths.

\subsection{Linear polarization data reduction}
\label{pred}

To extract the linear polarization information
at 86\,GHz we modelled the response of the two orthogonally
polarized linear feeds of the receivers A100 and B100 to a
partially linearly polarized source\footnote{
We assumed a negligible degree of circular polarization ($p_{\rm{C}}$),
which is null for the observed Galactic sources and typically
$< 1$\,\% at cm and mm wavelengths for extragalactic AGN
(Clemens Thum private communication; Homan \& Lister~\cite{Hom06}
and references therein). In particular, at 86\,GHz the measured
$p_{\rm{C}}$ of \object{0716+714} is $<0.4$\,\% (Clemens Thum
private communication)}.
In Appendix~\ref{appA} we show that, for our observing
set-up at the IRAM 30\,m telescope, these responses fulfil
the following equations:
   \begin{equation}
   \label{S_0}
   \frac {S_{\rm{A100}}(i) + S_{\rm{B100}}(i)}{2}=S_{0}(i) \,,
   \end{equation}

   \begin{equation}
   \label{pol1}
   S_{\rm{A100}}(i)-S_{0}(i)
   =\frac{1}{2} P(i) \cos{2\left(\alpha(i)+\frac{\pi}{2}\right)} \,
   \end{equation}
and
   \begin{equation}
   \label{pol2}
   S_{\rm{B100}}(i)-S_{0}(i)
   =\frac{1}{2} P(i) \cos{2(\alpha(i))} \,,
   \end{equation}
where $S_{\rm{A100}}(i)$ and $S_{\rm{B100}}(i)$ are,
respectively, the flux densities recorded by the receivers
A100 and B100 at the time of the $i^{\rm{th}}$ measurement,
$S_{0}(i)$ is the total flux density of the source,
$P(i)$ is its { linearly} polarized flux density and
$\alpha(i)$ is the angle between the horizontal axis of the
telescope's polarization reference plane (coincident with
the B100 polarization axis) and the direction of
polarization of the source at time $i$.
The angle $\alpha(i)$ is defined as
   \begin{equation}
   \label{alpha}
   \alpha(i)=\frac{\pi}{2}-\chi(i)+\eta(i)-\epsilon(i) \,,
   \end{equation}
where $\chi(i)$ is the astronomical electric vector
{ position} angle { (polarization angle hereafter)}
of the source (defined in the equatorial coordinate
system from north to east). $\eta(i)$ and $\epsilon(i)$
are, respectively, the parallactic and elevation angles
at the time of the $i^{\rm{th}}$ measurement (see
Appendix~\ref{appA} and Thum et al.~\cite{Thu00}).
We estimated the magnitudes of $P(i)$ and $\chi(i)$ by
performing independent fits on the equations (\ref{pol1})
and (\ref{pol2}) to the data.
It was necessary to assume that the source did not
vary in polarization during the time interval of each fit,
$5\,\rm{h}\aplt t \aplt 12$\,h.
To reduce the final uncertainty of the results,
the two independent estimates of $P(i)$ and $\chi(i)$ (from
the independent fits of (\ref{pol1}) and (\ref{pol2})) were
averaged together.

{ Thum et al. (\cite{Thu03}) reported that, for
polarimetric measurements with the A100 and B100 receivers
of the IRAM 30\,m telescope, the main contribution to the
instrumental polarization is produced by differences in
the relative calibration of both receivers.
As explained in \S~\ref{sred}, special care was taken to
remove these calibration differences {\it i}) by correcting
for the independent small pointing offsets
(typically $\aplt$ 3\hbox{$^{\prime\prime}$}) and {\it ii})
by correcting for the gain ratio affecting the A100 and B100
receivers.
Hence, the main component of the instrumental polarization
is initially removed from the data through our calibration
procedure.}

{ To test this and to quantify the amount of spurious
polarization still remaining in the data, we performed fits
of the normalized (to $S_{0}(i)$) versions of equations (\ref{pol1})
and (\ref{pol2}) to the data of the unpolarized calibrators
(\object{Mars}, \object{Uranus}, \object{W3\,OH}, \object{K3-50A}
and \object{NGC\,7027}) over the whole observing time range.
The result, $<p>=(0.2\pm0.9)\,\%$, provides a measurement
of the maximum instrumental linear polarization
affecting our polarization fits,
$p_{\rm{inst}}^{\rm{max}} = 1.1\,\%$, which we consider
negligible for the purposes of our data analysis.
The fact that $p_{\rm{inst}}^{\rm{max}}$ is of similar
magnitude than $\rm{rms}_{86}$ indicates that, as
for the total flux density measurements, our
polarization accuracy is most likely limited
by the statistical uncertainties induced by the
small and rapid atmospheric fluctuations.

Such uncertainties are computed from the linear
polarization fits.
In the case of \object{0716+714}, for time periods
$t \aplt 12$\,h, they lie in the range 1.2\,\%
to 3.4\,\% for $p$ and between $\sim$0.06\,Jy and
$\sim$0.12\,Jy for $P$ (see \S~\ref{pres}).
The additional error (a 5\,\% of the total flux
density) coming from the inaccuracy on the
absolute flux density scaling, has been added in
quadrature wherever it was necessary for the final
$P$ results; i.e. only for non-relative variability
results.
In such cases, we explicitly indicate that this
additional error was added.

The accuracy of our polarization angle
estimates is better tested on sources with large
linear polarization fraction, which should show
large modulation amplitudes in functions
(\ref{pol1}) and (\ref{pol2}), and hence should
allow for a more accurate measurement of $\chi$.
Among our observed sources, \object{0716+714} was the most
suitable one for such test, since it showed the
highest polarization degree
(see \S~\ref{res}) and was better time sampled.
The two independent fits of functions (\ref{pol1}) and
(\ref{pol2}) to the \object{0716+714} data, over ten different
time ranges of $t \aplt 12$\,h (\S~\ref{pres}),
showed an internal consistency of the $\chi$
estimates to a precision better than $0.5^{\circ}$.
This demonstrates the
accuracy of equations (\ref{pol1}), (\ref{pol2})
and (\ref{alpha}) and the good orthogonality
of the signals recorded by receivers A100 and B100.
However, the statistical uncertainties on the
determination of $\chi$ fail to be better than
$\chi_{\rm{inst}}^{\rm{max}}=1.1^{\circ}$ even for
the polarization fits to the \object{0716+714} data over
the whole observing time.
Also for \object{0716+714}, these uncertainties range
between $1.8^{\circ}$ and $9^{\circ}$ for the
ten time spans with $t \aplt 12$\,h
(\S~\ref{pres}).

Note that the accuracy of \emph{absolute} polarization
angle measurements obtained from A100 and B100
data is driven by the accuracy of the orientation
of their polarization axes (see Appendix~\ref{appA} and
Thum et al.~\cite{Thu00}).
Such orientations were calibrated with an accuracy
of $\sim 1^{\circ}$ (Clemens Thum, private
communication), which is therefore the better absolute
polarization angle accuracy achievable with the
A100 and B100 receivers.
A corresponding error of $1^{\circ}$ has been added
quadratically to the $\chi$ statistical uncertainties
of our measurements obtained for the whole observing
time range, but not to those computed for shorter time
spans -which are aimed at relative variability
analyses-.}
\\


\section{Results}
\label{res}

\begin{figure}
   \centering
   \includegraphics[bb=0 0 522 493,width=8.5cm,clip]{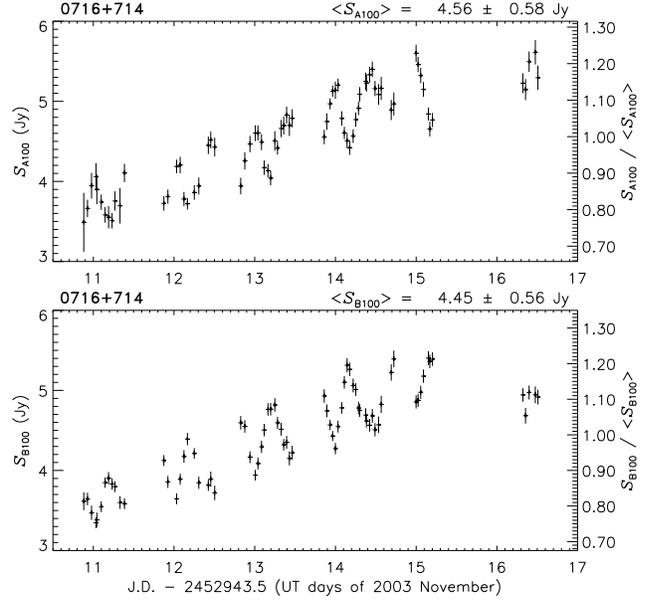}
      \caption{Calibrated measurements of $S_{\rm{A100}}$ and $S_{\rm{B100}}$
              for \object{0716+714} vs. time during 2003 November 10 to 16.
	      The sinusoidal pattern is due to source polarization
	      (see the text).}
   \label{sA100B100}
\end{figure}

In Fig.~\ref{sA100B100} we present plots of the resulting calibrated
responses of receivers A100 and B100 to the \object{0716+714} emission
during November 11 to 17. The curves show a remarkable increase of
$\sim1.5$\,Jy in $S_{\rm{A100}}$ and $S_{\rm{B100}}$ during
November 10 to 14 together with a large amplitude sinusoidal
modulation, which has a characteristic periodicity of 12\,h
and has a phase difference of $\pi$\,rad between the
$S_{\rm{A100}}$ and $S_{\rm{B100}}$ patterns.
The main contribution to this modulation { is explained by} 
the right-hand terms of expressions (\ref{pol1}) and
(\ref{pol2}), which are { significative} when the source is
highly linearly polarized. 
{ Following (\ref{pol1}) and (\ref{pol2}), the linear
polarization flux density of \object{0716+714} can be estimated
from the amplitudes of the modulations of the $S_{\rm{A100}}$ 
and $S_{\rm{B100}}$ curves, which lie in the range
$0.4\,Jy \aplt P \aplt 0.9\,Jy$.}
Hence, the curves in Fig. \ref{sA100B100}
represent the first evidence of a high degree of linear
polarization { ($10\,\%\aplt p \aplt 20\,\%$)}
in \object{0716+714} at 86\,GHz.
As far as we know, such { a high linear polarization}
was never observed before { in the integrated emission
of \object{0716+714} in any radio to mm band.}

\begin{figure}
   \centering
   \includegraphics[bb=0 0 531 736,width=8.5cm,clip]{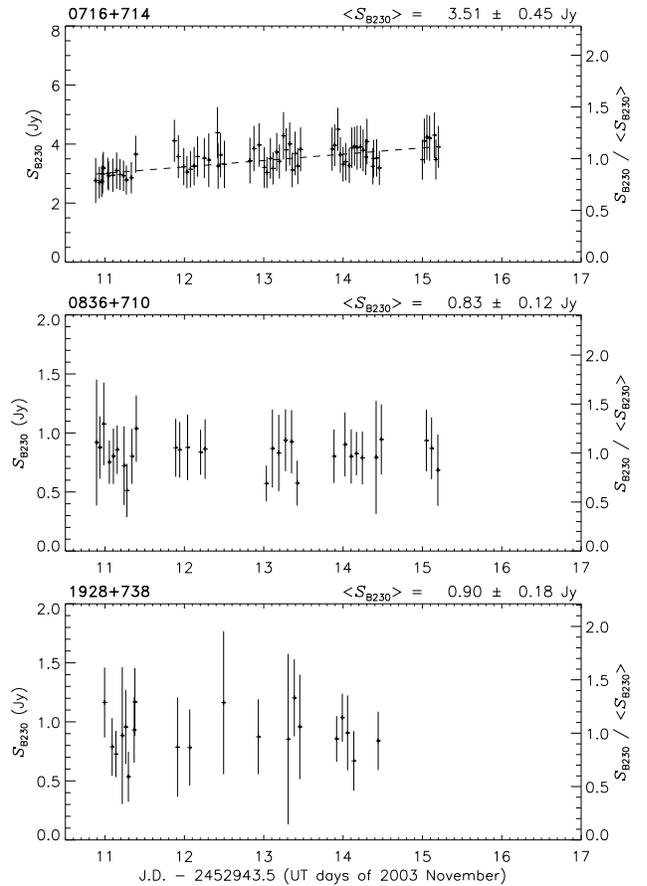}
      \caption{Calibrated measurements of $S_{\rm{B230}}$ for
              \object{0716+714}, \object{0836+710} and
	      \object{1928+738} during 2003 November 10 to 15.
	      The dashed line on the top inset represents the best
	      linear fit to the \object{0716+714} light curve.}
   \label{tfl1mm}
\end{figure}

In Fig. \ref{tfl1mm} we show the calibrated response of B230 to
the \object{0716+714} emission during November 10 to 15.
The larger measurement uncertainties at 229\,GHz,
which are typically $\sim 18$\,\%, do not allow us to detect
IDV with amplitude lower than this level.
Therefore, we are able to determine only the basic
emission properties at this frequency.
Since at 229\,GHz \object{0716+714} does not show the expected
12\,h periodic polarization modulation, then, within the
errors, the response of B230 is a valid measurement of the source
total flux density ($S_{229}$).

\subsection{Total flux density}
\label{sres}

\begin{figure}
   \centering
   \includegraphics[bb=0 0 541 745,width=8.5cm,clip]{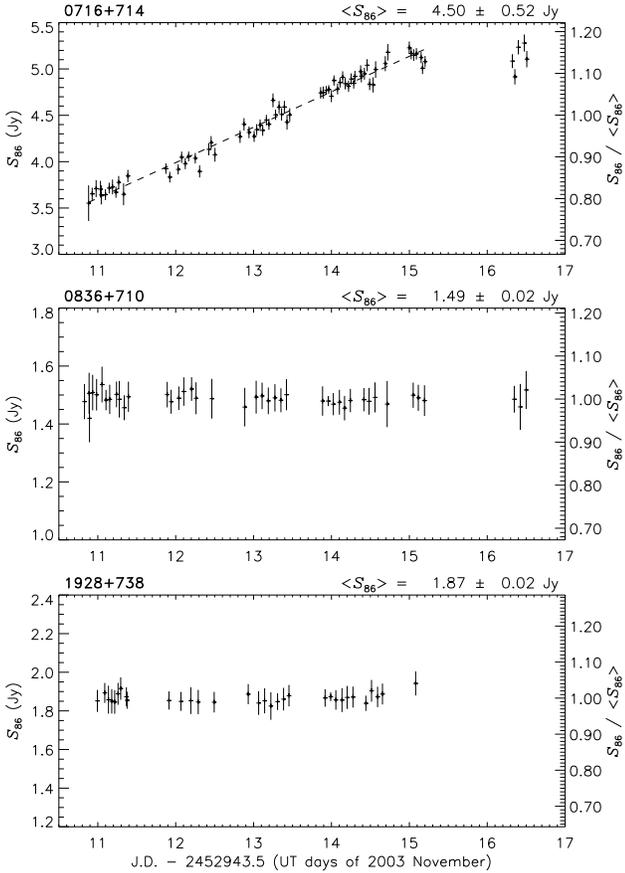}
      \caption{Calibrated 86\,GHz light curves of \object{0716+714},
              \object{0836+710} and \object{1928+738}
	      during 2003 November 10 to 16. The dashed line on the
	      top inset represents the best linear fit to the
	      \object{0716+714} flux density evolution during 2003 November
	      10 to 15.}
   \label{tfl3mm}
\end{figure}

Accurate measurements of the 86\,GHz flux densities ($S_{86}(i)$)
were obtained by averaging together $S_{\rm{A100}}(i)$ and
$S_{\rm{B100}}(i)$ (see Appendix~\ref{appA}).
The resulting evolution of $S_{\rm{86}}(i)$ during the time range
of our observations of \object{0716+714} is presented in Fig. \ref{tfl3mm}
together with the light curves of the non-IDV sources \object{0836+710}
and \object{1928+738}. The latter demonstrate the accuracy of our
calibration, which provided flux density measurements characterised
by an $rms_{86} \approx 1.2$\,\%. This represents an unprecedented
stability for 86\,GHz measurements with the IRAM 30\,m telescope.

The variability of \object{0716+714} at 86\,GHz appears to be dominated
by a monotonous linear increase from $<S_{86}>=(3.71\pm0.08)$\,Jy
during November 10 to 11 to $<S_{86}>=(5.13\pm0.86)$\,Jy during
November 15 (with $< >$ denoting averaging in time).
During the $\sim 5$\,h of data obtained on November 16, the
source displayed a similar mean flux density of
$<S_{86}>=(5.12\pm0.13)$\,Jy.
The linear fit of the light curve from days 10 to 15
(Fig. \ref{tfl3mm}) gives an increasing rate
($\dot{S}_{86}/<S_{86}>$) of $(8.58\pm0.14)$\,\%/day.
The fit performed to the 229\,GHz light curve gives
$\dot{S}_{229}/<S_{229}> =(6.2\pm1.7)$\,\%/day, which shows a
similar trend at both observing frequencies. This suggests that
the variability at both 86\,GHz and 229\,GHz { originates}
from the same region of the source and the same physical process.

To characterise the flux density variability of each
source we used the statistical formulation described in
Quirrenbach et al. (\cite{Qui00}) and Kraus et al.
(\cite{Kra03}).
In particular, we made use of the modulation index $m$, the
variability amplitude $Y$, the reduced chi-squared
$\chi_{r}^2$ and the structure function $D(\tau)$. The
corresponding definitions are summarised in Appendix~\ref{appB}.

The computed statistical parameters for \object{0716+714} and the secondary
calibrators are listed in Table \ref{sstat} for both the 86\,GHz and
the 229\,GHz data set.
The computation of the variability amplitude $Y$ was performed only
for those sources that passed the chi-squared test
(variable with significance level $>99.9$\,\%).
As can be seen in the Table, only \object{0716+714} showed such
significant variability (at 86\,GHz, but not at 229\,GHz due to the
larger uncertainties at that frequency).

Although \object{1642+690} formally did not pass this test, it displayed
larger values of $m$ and $\chi_{r}^{2}$ than the remaining
calibrators at both 86\,GHz and 229\,GHz (see Table \ref{sstat}),
suggesting { a} weaker variability.
In fact, its 86\,GHz light curve (frequency at which
the variability analysis is more accurate) is variable at
confidence level $> 95$\,\%.
Based on this evidence, we excluded \object{1642+690} as secondary
calibrator from the 86\,GHz and 229\,GHz data reduction.

\begin{table}
\caption[]{Statistical parameters characterising the variability
          of \object{0716+714} and the secondary calibrators during 2003
	  November 10 to 16.}
\begin{flushleft}
\small
\begin{tabular} {lcccccc}
\hline\noalign{\smallskip}
Source &  N & $<S>$ & $\sigma_{S}$ & $m$ & $\chi_{r}^{2}$ & $Y$\\
       &    &  [Jy] &    [Jy]      & [\%]&                &[\%] \\
\hline\noalign{\smallskip}
\multicolumn{7}{c}{$\nu=86$\,GHz,  $m_{0}=1.20$\,\%}\\
\hline\noalign{\smallskip}
\object{0212+735} & 33 & 1.101 & 0.016 &  1.45 &  0.3  &  ...  \\
\object{0633+734} & 25 & 0.942 & 0.018 &  1.90 &  0.4  &  ...  \\
\object{0716+714} & 74 & 4.502 & 0.516 & 11.46 & 55.3  & 34.19 \\
\object{0836+710} & 42 & 1.486 & 0.020 &  1.36 &  0.1  &  ...  \\
\object{1642+690}$^*$ & 24 & 1.044 & 0.060 &  5.73 &  1.6  &  ...  \\
\object{1803+784} & 32 & 1.334 & 0.016 &  1.19 &  0.1  &  ...  \\
\object{1928+738} & 32 & 1.866 & 0.024 &  1.31 &  0.1  &  ...  \\

\hline\noalign{\smallskip}
\multicolumn{7}{c}{$\nu=229$\,GHz,  $m_{0}=16.0$\,\%}\\
\hline\noalign{\smallskip}
\object{0212+735} & 15 & 0.712 & 0.094 & 13.14 & 0.1  & ... \\
\object{0633+734} & 15 & 0.621 & 0.085 & 13.72 & 0.1  & ... \\
\object{0716+714} & 68 & 3.506 & 0.448 & 12.77 & 0.5  & ... \\
\object{0836+710} & 31 & 0.830 & 0.123 & 14.77 & 0.3  & ... \\
\object{1642+690}$^*$ &  8 & 0.744 & 0.254 & 34.17 & 0.5  & ... \\
\object{1803+784} & 20 & 0.940 & 0.235 & 25.04 & 0.8  & ... \\
\object{1928+738} & 20 & 0.904 & 0.177 & 19.54 & 0.4  & ... \\

\noalign{\smallskip}
\hline
\end{tabular}
\end{flushleft}
$^*$Although this source did not pass our chi-squared test,
it displays much larger $m$ and $\chi_{r}^{2}$ than the
remaining secondary calibrators, suggesting a { higher}
variability.

Note: $Y$ computation was only performed for \object{0716+714}
at 86\,GHz, the only case for which
the variability chi-squared test was passed.
\label{sstat}
\end{table}

\begin{figure}
   \centering
   \includegraphics[bb=50 50 707 546,width=8.5cm,clip]{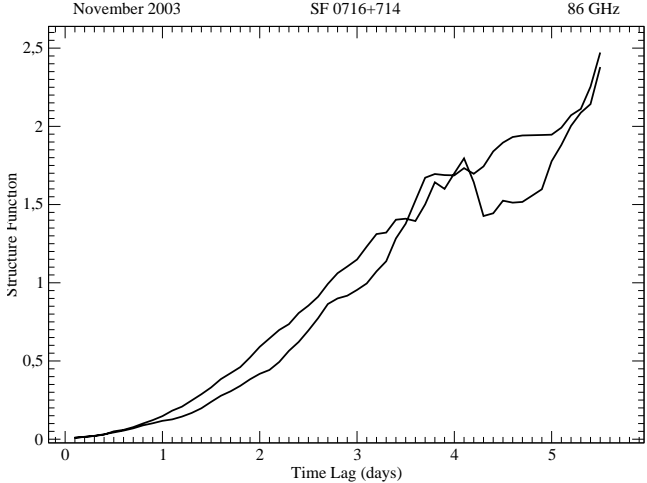}
      \caption{$D(\tau)$ function for \object{0716+714} at 86\,GHz. Since the
               data are unevenly sampled, we derived $D(\tau)$ using
	       an interpolation procedure. A minimal interpolation
	       interval of $\sim 2.5$\,h (3 to 5 times larger than the
	       typical $S_{86}$ sampling time-range) was chosen. To
	       assess the errors caused by the interpolation, $D(\tau)$
	       was derived twice, first by starting at the beginning of
	       the time series and second by proceeding backwards.}
   \label{sf3mm}
\end{figure}

The structure function $D(\tau)$ for \object{0716+714} at 86\,GHz is
shown in Fig. \ref{sf3mm}. The shape of this function provides
the variability type classification of the observed sources
(Heeschen et al.~\cite{Hee87}; Kraus et al.~\cite{Kra03}).
The lack of pronounced maxima and of a saturation point in $D(\tau)$
on time lags shorter than two days allows us to classify
\object{0716+714} as an IDV source of type I during the time range
of our observations at 86\,GHz.
The local maximum appearing at a time lag of $\sim4$\,days
characterises the time-scale of the monotonic increase of the
flux density $S_{86}$.
In agreement with this, and based on the same data set,
Fuhrmann et al. (\cite{Fuh06}) report, from a more sophisticated
estimate, a time-scale of $3.83^{+0.14}_{-0.15}$\,days.

\subsection{Polarization}
\label{pres}

\begin{figure*}
   \centering
   \includegraphics[bb=0 0 758 726,width=17.5cm,clip]{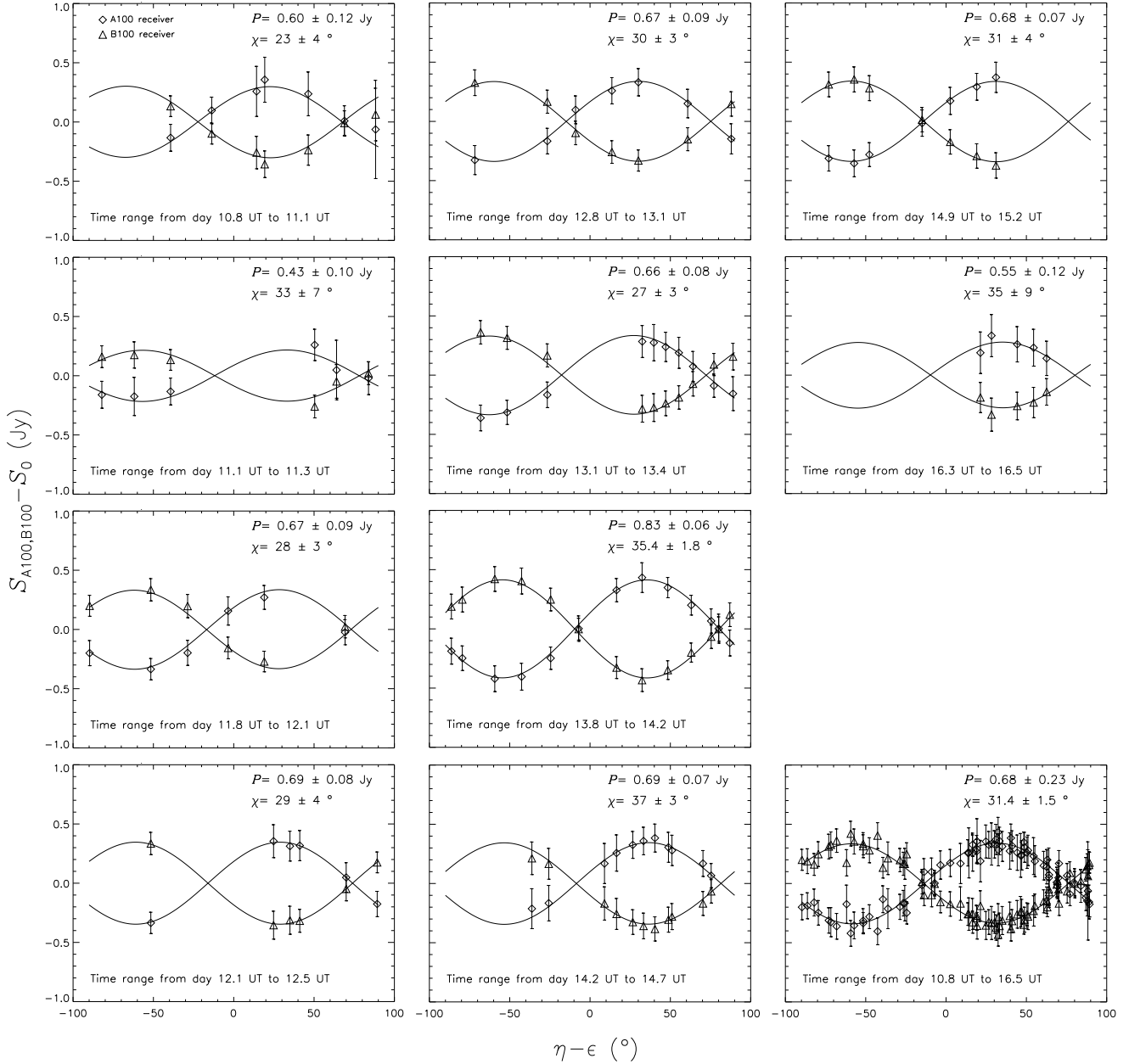}
      \caption{Plots of the fits of functions (\ref{pol1}) and
      (\ref{pol2}) to the 86\,GHz data of \object{0716+714}. Each
      plot represents a time bin $\aplt 12$\,h of observing time
      (labelled within each of them) with a minimum of ten data
      points. The time spans, ordered from top to bottom and
      from left to right, cover the whole observing time.
      The whole data set has been plotted together in the
      right-bottom plot. The sinusoidal lines represent the
      independent non-linear least squares fits of functions
      (\ref{pol1}) and (\ref{pol2}).
      The resulting averaged $P$ and $\chi$ variables are
      given within each corresponding plot.
      Errors were { originally}computed as 1-$\sigma$
      uncertainties from the fits.
     { For the right-bottom plot, the errors from
      the absolute total flux density and polarization angle
      uncertainties (see \S~\ref{obsred}) have been taken
      int account (added in quadrature)}.}
   \label{polfit}
\end{figure*}

\begin{figure*}
   \centering
   \includegraphics[bb=0 0 568 277,width=17.5cm,clip]{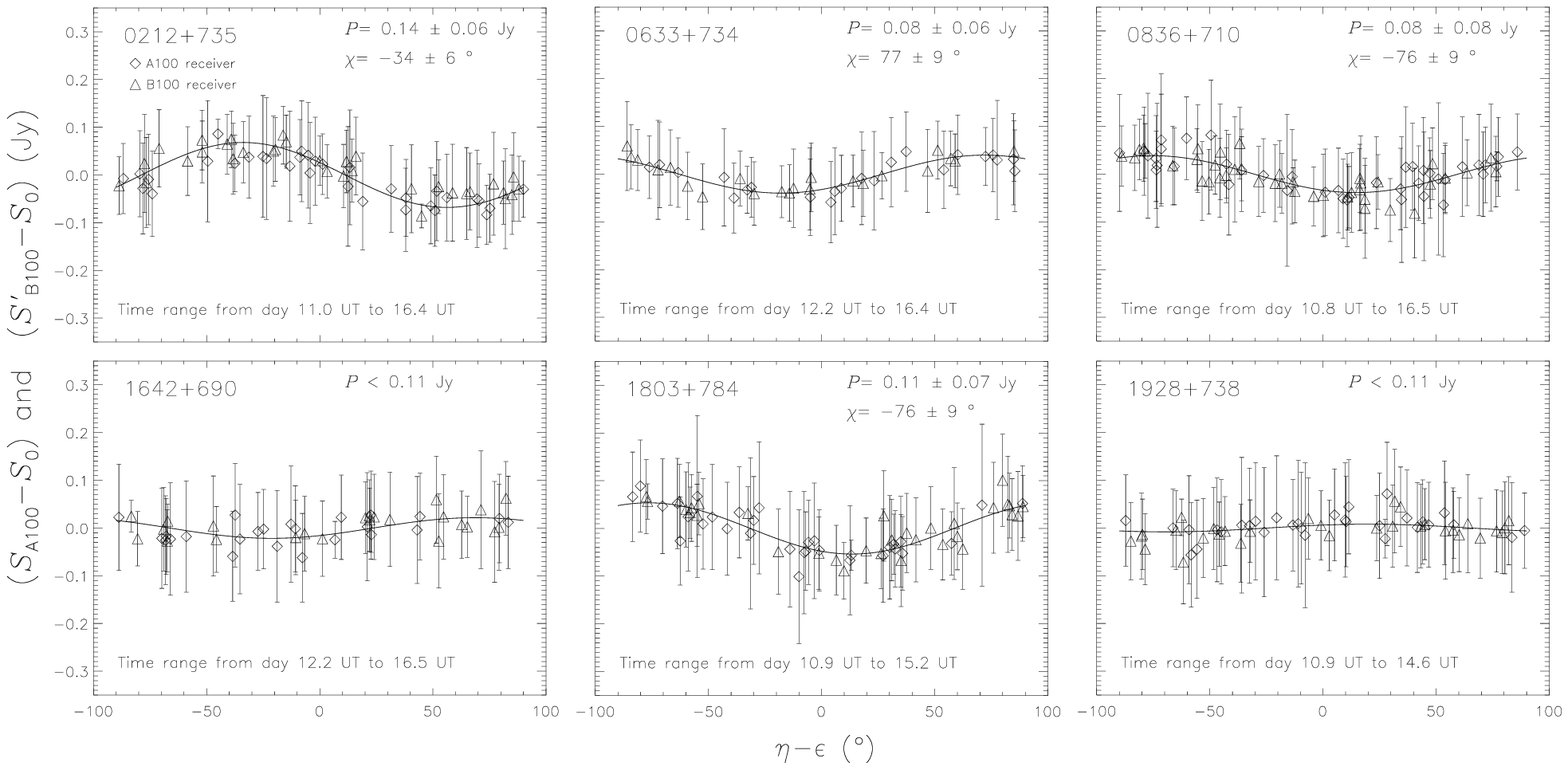}
      \caption{86\,GHz polarization patterns of the secondary
      calibrators. For the sake of clarity, a shift
      of $\pi$\,rad was added to function (\ref{pol2}) to match
      function (\ref{pol1}), i.e.
      $S'_{\rm{B100}}(i)-S_{0}(i)=1/2 P(i) \cos{2(\alpha(i)+\pi/2)}$
      (see \S~\ref{pred}).
      For each source, the data from the whole observing time-range
      is presented.
      The sinusoidal lines represent the non-linear least squares
      fits to the data.
      The resulting averaged $P$ and $\chi$ variables are
      given within each corresponding plot.
      The errors { presented on each plot} were computed
      { by adding in quadrature the} 1-$\sigma$
      uncertainties from the fits and the { errors from
      the absolute total flux density and polarization angle
      uncertainties (see \S~\ref{obsred})}.}
   \label{polfitcal}
\end{figure*}

The 86\,GHz polarization estimates were obtained by fitting
the functions (\ref{pol1}) and (\ref{pol2}), which relate
directly the amplitudes and phases of such polarization patterns
to the linearly polarized flux density $P$ and electric vector
{ position} angle $\chi$ of the observed source, respectively.
The fits performed for \object{0716+714} for $\aplt 12$\,h time bins (with
a minimum of ten points per fit) are shown in Fig. \ref{polfit}.
This time binning was chosen as the optimum to provide more than
one $P$ and $\chi$ measurement per day and the lowest
possible uncertainties in their estimate.
Polarization fits for { $\aplt \rm{ 24}$\,h time bins and for}
the entire time range of the observations were also performed for
{ both} \object{0716+714} { and} the secondary calibrators
(Fig.~\ref{polfitcal}, Table~\ref{pfitall}
{ and Table~\ref{pstat}}).

\begin{table}
\caption[]{Average 86\,GHz polarization properties of \object{0716+714}
          { and} the secondary calibrators during 2003 November 10
	  to 16.}
\begin{flushleft}
\small
\begin{tabular} {lccc}
\hline\noalign{\smallskip}
Source &  $<P>$\,[Jy] & $<p>$\,[\%] & $<\chi>$\,[$^{\circ}$]  \\
\hline\noalign{\smallskip}
\object{0212+735}      &   $  0.14\pm$0.06     &   $   12\pm2    $   &   -34$\pm$6  \\
\object{0633+734}      &   $  0.08\pm$0.06     &   $    8\pm3    $   &   $   77\pm$9     \\
\object{0716+714}      &   $  0.68\pm$0.23     &   $  15.0\pm1.8 $   &   $ 31.4\pm$1.5   \\
\object{0836+710}      &   $  0.08\pm$0.08     &   $   5.3\pm1.6 $   &   -76$\pm$9  \\
\object{1642+690}      &   { 0.04}$_{\rm{ -0.04}}^{\rm{ +0.07}}$      &   $    4\pm4    $   &   $   66\pm$18     \\
\object{1803+784}      &   $  0.11\pm$0.07     &   $    8\pm2    $   &   $  -76\pm$9  \\
\object{1928+738}      &   { 0.02}$_{\rm{ -0.02}}^{\rm{ +0.09}}$     &    { 1.1}$_{\rm{ -1.1}}^{\rm{ +1.6}}$      &   $      ...    $  \\

\noalign{\smallskip}
\hline
\end{tabular}
\end{flushleft}
Note: These results have been obtained, for each source,
from the polarization fits over the entire time range of
the observations. The errors were { originally}
computed as 1-$\sigma$ uncertainties of the polarization
fits. { The errors of the $<P>$ and $<\chi>$ measurements
include those from the absolute total flux density and
polarization angle uncertainties, respectively (see \S~\ref{obsred}).}
\label{pfitall}
\end{table}

The fits presented in Fig. \ref{polfit} confirm the
previously mentioned  high polarization level of \object{0716+714}
at 86\,GHz.
The average polarization properties of the source
(during our observations) were $<p>=(15.0\pm1.8)$\,\%, or
$<P>=(0.68\pm\rm{ 0.23})$\,Jy, and
$<\chi>=31.4^{\circ}\pm\rm{ 1.5}^{\circ}$ (Table \ref{pfitall}),
{ where the errors of $<P>$ and $<\chi>$ take into account
those from the absolute total flux density and polarization
angle uncertainties, respectively (see \S~\ref{obsred})}.
Such a high polarization degree is rather unusual for AGN at
centimetre wavelengths, for which $p>10$\,\% is rarely
observed even during flaring states (e.g. Aller et al.
\cite{All99}) and typically 1\,\%$\aplt p\aplt6$\,\%.
Synchrotron self-absorption, usually dominating at low
centimetre radio frequencies, can reduce
$p$ (e.g. Pacholczyk~\cite{Pac70}). However, low values
of polarization degree are observed at centimetre
wavelengths even for optically thin sources.
Numerical simulations show that, in order to explain
these low levels of polarization in compact radio sources,
a large fraction of the magnetic field (typically
50\,\%-70\,\%) should be disordered in direction
(e.g. G\'omez, Alberdi \& Marcaide~\cite{Gom93}).
The 15\,\% polarization degree detected in \object{0716+714} might
indicate a high level of magnetic field alignment
in the region from which the 86\,GHz emission originated.

In addition, one of the six AGN calibrators observed during
our observations, \object{0212+735}, also displays a
polarization degree $p>10$\,\% and \object{0633+734} and
\object{1803+784} show $p\approx8$\,\%.
These values are still high when compared with the typically
observed polarization degrees at centimetre wavelengths.
The fact that, at 86\,GHz, more than 50\,\% of the observed
AGN display larger polarization degrees than those typically
measured at centimetre wavelengths might suggest that the
polarization degree in AGN increases with frequency.
This would indicate higher ordering of the magnetic
fields in the deeper regions of their jets observed in the
millimetre bands respect to centimetre wavelengths.
However, it is clear that our sampled set of AGN is not
statistically significant enough to strongly support
this hypothesis.
{ The} millimetre polarization survey { performed
by Thum et al.~(in prep.)} over { more than 70} AGN
{ during the summer of 2006 will certainly} improve the
statistics.

\begin{figure}
   \centering
   \includegraphics[bb=0 0 665 500,width=8.5cm,clip]{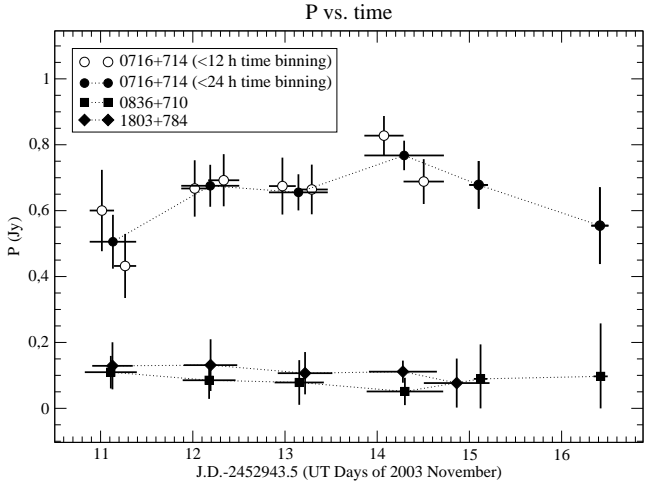}
      \caption{86\,GHz $P$ curves for \object{0716+714}, \object{0836+710} and
               \object{1803+784}. { Time bins of $<24$\,h were used for
	       all the sources. For \object{0716+714} the results for time
	       bins $\aplt 12$\,h are also presented.} Horizontal bars
	       symbolise the time range over which each estimate of $P$ was
	       computed.}
   \label{pcurves}
\end{figure}

\begin{figure}
   \centering
   \includegraphics[bb=0 0 665 504,width=8.5cm,clip]{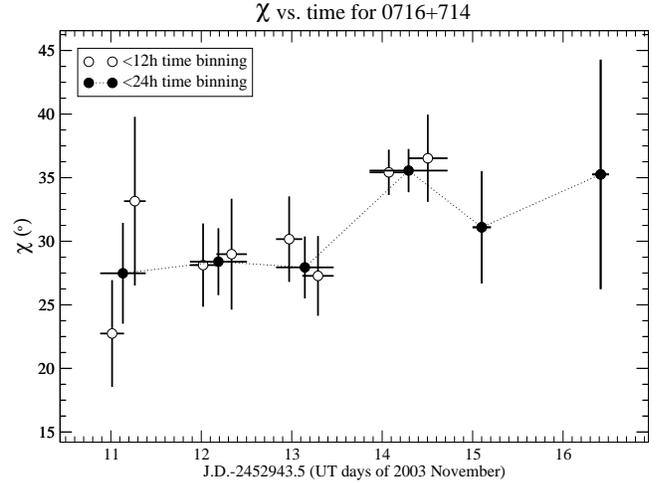}
      \caption{86\,GHz polarization angle vs. time for \object{0716+714}.
               { The results obtained for time bins of both
	       $<24$\,h and $\aplt 12$\,h are presented.}
               Horizontal bars symbolise the time range over which each
	       $\chi$ estimate was computed. For the sake of clarity,
	       the $\chi$ time series of the calibrators (which are all
	       constant within their corresponding errors) have not been
	       plotted due to their large uncertainties, typically of
	       15$^{\circ}$-20$^{\circ}$.}
   \label{chicurves}
\end{figure}

Inspection of Fig.~\ref{polfit} indicates time-dependent
changes in both the amplitude and the phase of the fitted
functions, suggesting that the linear polarization
of \object{0716+714} varied along our observations.
The $P$ estimates, for both \object{0716+714} and the secondary
calibrators, are presented as a function of time
in Fig.~\ref{pcurves}, while the $\chi$ evolution
for \object{0716+714} only is shown in Fig.~\ref{chicurves}.
Due to the much lower polarization degree of the calibrators,
their polarization properties were optimally fitted with
time bins of $< \rm{ 24}$\,h.
This provided one polarization measurement per day to test the
stability of the secondary calibrators and it also decreased
the fitting errors compared to shorter time-binnings.
{ To allow for a better comparison with the calibrators,
the \object{0716+714} data were also fitted using a
$< \rm{ 24}$\,h time binning (Fig.~\ref{pcurves},
Fig.~\ref{chicurves} and Table~\ref{pstat}).}

While the secondary calibrators remained stable in polarization
within the errors, \object{0716+714} displayed polarization
variability during November 11 and 14.
To quantify this variability, we performed
a statistical analysis similar to the one presented in \S
\ref{sres}.
Due to the much weaker polarization of the secondary calibrators
compared to \object{0716+714}, the errors of their $P$ and especially
their $\chi$ estimates are much larger than for \object{0716+714}.
Hence, { in principle, it would not be} reasonable to calculate
$m_{P}$, $m_{\chi}$, $Y_{P}$ and $Y_{\chi}$ since they do
not take into account the uncertainties on the measurements
and they would not provide reliable variability results.
However, the mean, the standard deviation and the reduced chi-squares
of both the $P$ and $\chi$ distributions can still provide useful
statistical information.
These magnitudes are summarised in Table \ref{pstat}.
None of the sources analysed passed the chi-squared test
neither for $P$ nor for $\chi$. However, \object{0716+714}
displayed relatively large values of $\chi_{r,P}^{2}=\rm{ 2.4}$ and
$\chi_{r,\chi}^{2}=\rm{ 2.2}$, with corresponding variability
significance levels $>\rm{ 95}$\,\%.
{ Hence, for the particular case of \object{0716+714} and given the
relatively large significance of its polarization variability,
it makes sense to compute both $m_{P}=15$\,\% and
$m_{\chi} \equiv \sigma_{\chi}=6^{\circ}$,
which are attributed to the source behaviour during the first
four observing days (Figs.~\ref{pcurves} and \ref{chicurves}).}

\begin{figure}
   \centering
   \includegraphics[bb=0 0 666 499,width=8.5cm,clip]{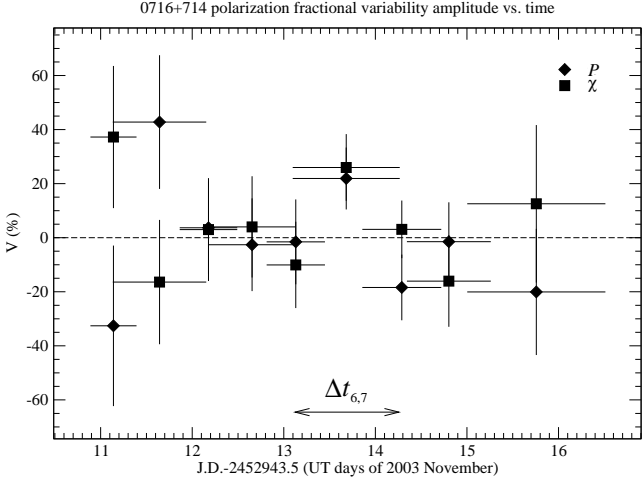}
      \caption{Amplitude of fractional polarization variability,
               as a function of time. The vertical bars represent
	       the $V_{P}$ and $V_{\chi}$ propagated errors whereas
	       the horizontal bars symbolise the maximum possible
	       time range over which the variability took place.
	       See the text for definitions.}
   \label{V}
\end{figure}

The amount of $P$ and $\chi$ variability of \object{0716+714} between
each pair of adjacent measurements of { these} variables was
characterised through their fractional variability amplitudes:

   \begin{equation}
   \label{Vp}
   V_{P}(\Delta t_{i-1,i})=\frac
   {2 (P(\Delta t_{i})-P(\Delta t_{i-1}))}
   {P(\Delta t_{i})+P(\Delta t_{i-1})}  \,
   \end{equation}
and
   \begin{equation}
   \label{Vchi}
   V_{\chi}(\Delta t_{i-1,i})=\frac
   {2 (\chi(\Delta t_{i})-\chi(\Delta t_{i-1}))}
   {\chi(\Delta t_{i})+\chi(\Delta t_{i-1})}\,,
   \end{equation}
where $\Delta t_{i}$ (with $i=1,2,...,10$ { for the \object{0716+714}
polarization measurements performed with a time binning
$\aplt \rm{\rm 12}$\,h}) denotes for the
interval over which each one of the polarization fits
was performed and $\Delta t_{i-1,i}$ symbolises the time
interval between each pair of adjacent polarization measurements.
We considered that an intra-day polarization variation
event is significant when either $V_{P}$ or $V_{\chi}$
is larger than { its} corresponding uncertainties
computed by propagating the errors of $P$ and $\chi$.
There { is} only one time interval ($\Delta t_{6,7}$) for which
this condition is fulfilled. $\Delta t_{6,7}$ is a time
interval of $\sim 24$\,h on November 13{-14}, for which both
$V_{P}(\Delta t_{6,7})=(22\pm11)$\,\% and
$V_{\chi}(\Delta t_{6,7})=(26\pm12)$\,\% show a $2\sigma$
variation in $P$ and $\chi$, respectively (see Fig. \ref{V}).
Hence, this simple analysis provided indications of IDV on the
polarized emission of \object{0716+714} during our observations.
Nevertheless, this statement should not be taken as a
strong evidence { of polarization IDV}, as no $3\sigma$
IDV event has been detected.
A future dedicated millimetre monitoring campaign, optimised
for high time sampling and high polarization sensitivity,
will be necessary to confirm and clarify whether
\object{0716+714} (or any other AGN) is variable in polarization on
time-scales of $\aplt 24$\,h.

\begin{table}
\caption[]{Statistical parameters characterising the 86\,GHz
          polarization variability of \object{0716+714} and the secondary
	  calibrators, but \object{1928+738}, during 2003 November 10--16.}
\begin{flushleft}
\small
\begin{tabular} {lccccccc}
\hline\noalign{\smallskip}
Source &  N & $<P>$ & $\sigma_{P}$ & $\chi_{r,P}^{2}$ &    $<\chi>$    & $\sigma_{\chi}$ & $\chi_{r,\chi}^{2}$ \\
       &    &  [Jy] &    [Jy]      &                  & [$^{\circ}$] &  [$^{\circ}$]   &                     \\
\hline\noalign{\smallskip}
\object{0212+735}     &  5 & 0.122 & 0.024 & 0.3 & -33.1 &  6.0 & 0.2\\
\object{0633+734}     &  5 & 0.092 & 0.021 & 0.1 &  69.3 &  8.5 & 0.3\\
\object{0716+714}$^a$     & 10 & 0.648 & 0.103 & 1.8 &  30.9 &  4.3 & 1.7\\
 \object{0716+714}$^b$     &   6 &  0.639 &  0.095 &  2.4 &   31.0 &   3.3 &  2.2\\
\object{0836+710}     &  6 & 0.085 & 0.020 & 0.2 & -76.3 &  8.5 & 0.1\\
\object{1642+690}     &  5 & 0.064 & 0.023 & 0.1 &  54.6 & 17.6 & 0.3\\
\object{1803+784}     &  5 & 0.111 & 0.022 & 0.1 &  76.4 &  4.5 & 1.2\\
\noalign{\smallskip}
\hline
\end{tabular}
\end{flushleft}
{ $^a$\object{0716+714} results for $\aplt 12$\,h.}

{ $^b$\object{0716+714} results for $<24$\,h are also presented
to allow direct comparison with the secondary calibrators.}

Note~1: $<P>$ and $<\chi>$ in this Table, which are
based on $P$ and $\chi$ fittings over
$<24$\,h time bins, are less accurate than those
from Table \ref{pfitall}, which where fitted over
the whole time span of the observations.

Note~2: The weak linearly polarized flux density of \object{1928+738}
(see Table~\ref{pfitall}) { does} not allow us to analyse its
variability.
\label{pstat}
\end{table}


\section{Discussion}

\subsection{Spectrum}

{ The millimetre data presented here contributed to the
assembling of the broad-band spectrum (from radio to
$\gamma$-rays) of \object{0716+714} during November 10 to 16,
2003 presented by Ostorero et al.~(\cite{Ost06}).
The spectrum was inverted} between the cm- and mm-bands.
Its turnover frequency ($\nu_{m}$, at which the synchrotron
spectrum peaks) was located at $\nu_{m} \approx 86$\,GHz.
Between 86\,GHz and 667\,GHz the spectrum can be fitted by a single
power law with spectral index $\alpha \approx -0.3$ (where $S_{\nu}
\varpropto \nu^{\alpha}$ and $\nu$ denotes the observing
frequencies within this spectral range).

\begin{figure}
   \centering
   \includegraphics[bb=25 8 669 504,width=8.5cm,clip]{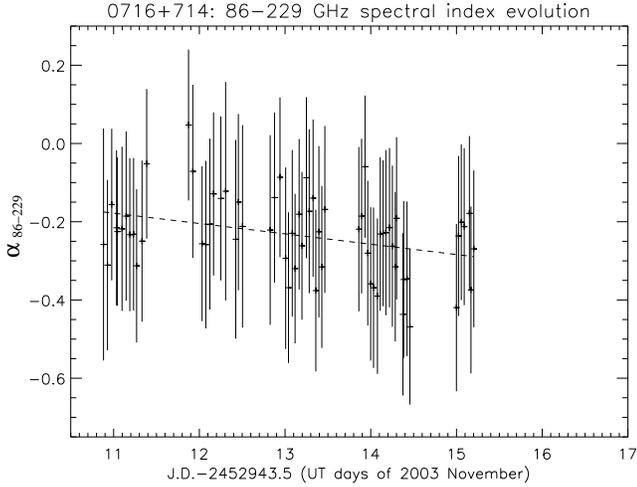}
      \caption{Evolution of the spectral index $\alpha_{86-229}$ in
               \object{0716+714}.
	       { The absolute total flux density calibration
	       errors of $S_{86}$ and $S_{229}$ (see \S~\ref{obsred})
	       were taken into account to compute those of $\alpha_{86-229}$.}
	       The dashed line represents the best linear
	       fit to the data, which is consistent with a
	       pattern with a rate $\dot{\alpha}_{86-229}=(\rm{ -0.03}\pm
	       \rm{ 0.02})\,\rm{d}^{-1}$ and average
	       $\bar{\alpha}_{86-229}=-0.23 \pm 0.10$.}
   \label{spev}
\end{figure}

At 86\,GHz, the total flux density of the source (\S \ref{sres})
ranged from $S_{m}=3.55\pm0.19$\,Jy at the 10.88 U.T. days of
November 2003 to $S_{m}= 5.23\pm0.07$\,Jy at the 15.00 U.T. days
of November 2003.
The larger total flux density of the source at 86\,GHz
compared to that at 229\,GHz, (Figs. \ref{tfl1mm} and \ref{tfl3mm})
shows that the source synchrotron spectrum was optically thin
between these two frequencies during the whole time range of
our observations.
To { attempt to} characterise the source spectral evolution, we have
computed the spectral index for each pair of $S_{86}(i)$ and
$S_{229}(i)$, where $i$ is the index defining each one of the
simultaneous 86\,GHz and 229\,GHz measurements.
The resulting $\alpha_{86-229}$ evolution (shown in Fig.
\ref{spev}) is consistent with a decreasing pattern (spectral
softening) with constant rate $\dot{\alpha}_{86-229}=(\rm{ -0.03}\pm
\rm{ 0.02})\,\rm{d}^{-1}$ and average $\bar{\alpha}_{86-229}=-0.23 \pm
0.10$ from the 10.88 to the 15.00 U.T. days of November 2003.
{ However, the computed $\dot{\alpha}_{86-229}$ is not
statistically significant and the possible spectral change of
\object{0716+714} needs to be tested through more accurate spectral
index data.}

\subsection{Intrinsic variability}
\label{var}

{ Given the known frequency dependence for weak ISS,}
it is rather { unlikely} that the $Y=34$\,\% amplitude
inter-day variability reported { at 86\,GHz} is caused by
{ this extrinsic effect}.
As outlined in \S \ref{int}, such large variability in the
mm-bands can not be reproduced by the standard ISS models
{ (e.g. Beckert et al.~\cite{Bec02}).}
Fuhrmann et al. (\cite{Fuh06}) perform a { more detailed}
variability study { combining all the available
simultaneous data from 5\,GHz to 667\,GHz}, and show
that { the} inter-day variations shown { in this paper}
contradict the predictions of such models.
{ We should hence attribute the large amplitude 86\,GHz
variability of \object{0716+714} to intrinsic causes.}

{ Hence, the high amplitude inter-day variability
reported by Ostorero et al.~(\cite{Ost06}) at 32\,GHz and
37\,GHz, which well { matches} with the millimetre
behaviour shown in this paper, was most likely produced
in the source itself and not by ISS.
Ostorero et al. (\cite{Ost06}) also report on the rapid IDV
in the optical range observed during the time range of our
IRAM 30\,m observations.
Although \object{0716+714} did not show evidence of
{ simultaneous radio or mm and optical IDV-time-scale
transitions}, as during previous monitoring campaigns
(Quirrenbach et al.~\cite{Qui91}), this can not be
taken as an argument in favour of the extrinsic origin of
the variability.
It indicates only that, during our observations, the
radio-mm emission was radiated from a different region,
or through a different mechanism, than that producing the
optical emission.}

\subsection{Brightness temperature}
\label{Tb}

Under the assumption that the main component of the
mm-variability is intrinsic to the source and via the causality
argument, it is possible to estimate some of the physical conditions
governing the source.
This can be performed by modelling it as a relativistically moving
homogeneous sphere (e.g. Marscher et al.~\cite{Mar79} and
Ghisellini et al.~\cite{Ghi93}).
Under these assumptions, the { intrinsic} brightness temperature
of the emitting region at { the turnover frequency} $\nu_{m}$
(at which the spectral index is zero, by definition) can be computed
from its variability time scale as:
   \begin{equation}
   \label{TB}
   T_{\rm{B}} = 1.77 \times 10^{12} \frac {S_{m}} {\nu_{m}^{2} \theta_{v}^{2}}
              {\left(\frac {1+z} {\delta}\right)}\, \rm{K},
   \end{equation}
where { $S_{m}$ is the flux density of the source at the
turnover frequency in Jy and $\nu_{m}$ is in GHz.}
$z$ is the redshift of the source, $\delta$ is an equivalent
Doppler factor and
   \begin{equation}
   \label{theta}
   \theta_{v} \aplt 3.56 \times 10^{-4} t_{v} {d_{\rm L}}^{-1}
              (1+z) \delta \, \rm{mas}
   \end{equation}
is an estimate of its angular diameter (Marscher et al.~\cite{Mar79})
from light-travel time arguments in the absence of coherence.
$t_{v} \equiv \mid {\rm d} {\rm ln} S_{\nu} / {\rm d}t \mid ^{-1} =
(<S_{\nu}>/\Delta S_{\nu}) \Delta t$ is the variability time scale
in days and $d_{\rm L}$ is the source luminosity distance in Gpc.
In this paper, we adopt
$z=0.3$ (the lower limit for the redshift of \object{0716+714}, see e.g. \S
\ref{int} and Wagner et al.~\cite{Wag96}), a Hubble constant
$H_{0}=72$\,km\,s$^{-1}$\,Mpc$^{-1}$ and a
Friedmann- Robertson-Walker cosmology with $\Omega_{m}=0.3$ and
$\Omega_{\Lambda}=0.7$. Within these assumptions, $d_{\rm L}=1.51$\,Gpc.

Our 86\,GHz observations provide us the values of $S_{m}=5.23$\,Jy,
$<S_{m}>=4.50$\,Jy, $\Delta S_{m}=1.68$\,Jy,
$\Delta t=4.12$\,days { and $t_{v}=\rm{{ 11.1}}$\,days}.
The only remaining variable in expression (\ref{TB}) which can not
be directly observed is $\delta$. By assuming
a non-moving source, expression (\ref{TB}) gives { us the
{\it apparent} brightness temperature}
$T_{\rm{B}}^{\rm{app}} \equiv T_{\rm{B}}(\delta=1)> 1.4 \times 10^{14}$\,K at $z=0.3$,
which exceeds by more than two orders of magnitude the IC-limit
$T_{\rm{B,IC}}^{\rm{lim}} \approx 10^{12}$\,K (Kellermann \& Pauliny-Toth
\cite{Kel69}). Hence, $T_{\rm{B,IC}}^{\rm{lim}}$ sets a lower limit to the
Doppler factor of the emitting region, which should then be
$\delta_{T_{\rm{B}}} > 5.2$. Note that a more accurate limit for $T_{\rm{B,IC}}^{\rm{lim}}$
(of $\sim 3 \times 10^{11}$\,K) was computed by Readhead~\cite{Rea94}
(see also Kellermann~\cite{Kel03}). To avoid the violation of the
latter, $\delta_{T_{\rm{B}}} > 7.8$ is required.

\subsection{Angular size}
\label{size}

The latter constraint ($\delta_{T_{\rm{B}}} > 7.8$), together with
expression (\ref{theta}), allows us to estimate an upper limit for
the size of the emitting region { responsible for the variability} 
during our observations, which should be 
$\theta_{T_{\rm{B}}} < 0.03$\,mas.
Obviously, $\theta_{T_{\rm{B}}}$ is also an upper limit for the
maximum size of { such region} obtained from $\delta_{T_{\rm{B}}}>5.2$.

{ It should be stressed that $\theta_{T_{\rm{B}}}$ does not account
for the size of the whole source, but only for the region responsible 
for the rapid variability, which in the case of \object{0716+714} has been
associated with the VLBI core component of its relativistic 
jet (Bach et al.~\cite{Bac06}).}

An independent constraint on the size of the { variability
region} can be obtained from the 86\,GHz VLBI observations of 
\object{0716+714} performed by Bach et al.~(\cite{Bac06}) in April 2003.
A Gaussian model fit of their resulting image gives a FWHM size 
of the { 86\,GHz VLBI core} of 0.014\,mas, which is much lower
than the minimum beam size of { VLBI} observations { at
such frequency} ($\sim 0.04$\,mas; e.g. Agudo et al.~\cite{Agu05}).
This gives us $\theta_{VLBI}<0.04$\,mas, which is in agreement 
with our { estimate of $\theta_{T_{\rm{B}}}$}.

We will hence assume hereafter that the true angular size of the
{ rapid variability region} is 
$\theta_{v} = \theta_{T_{\rm{B}}} < 0.03\,\rm{mas}$.

\subsection{Inverse Compton Doppler factor}
\label{dopIC}

It is also possible to compare the expected and the observed
first order self-Compton $\gamma$-ray flux density to constrain the
{\it IC Doppler factor} ($\delta_{\rm{IC}}$) of the source
(Marscher~\cite{Mar83} and Ghisellini et al.
\cite{Ghi93}). $\delta_{\rm{IC}}$ is defined as:

   \begin{equation}
   \label{dopx}
   \delta_{\rm{IC}} = f ( \alpha ) S_{m}
              \left[ \frac {\rm{ln}({\nu_{2}/\nu_{m}})
	      {\nu_{\gamma}}^{\alpha}}
	      { S_{\gamma} { \theta_{v} }^{ ( 6 - 4 \alpha ) }
	      { \nu_{m} }^{ ( 5 - 3 \alpha ) } }
	      \right]^{ 1 / ( 4 - 2 \alpha ) } ( 1 + z ) \,,
   \end{equation}
where { $\alpha$ is the spectral index of the optically thin
synchrotron spectrum ($\alpha=-0.3$ has been adopted here)},
$f(\alpha)=0.14 -0.08 \alpha$ (Ghisellini et al.
\cite{Ghi93} and references therein), $\nu_{2}$ is the upper
frequency cut-off of the synchrotron spectrum in GHz, $\nu_{\gamma}$
is the $\gamma$-ray { energy} in keV and $S_{\gamma}$ is the
measured $\gamma$-ray flux at $\nu_{\gamma}$ in Jy.
$\nu_{2}$, in GHz, is the most poorly
known variable among those in equation (\ref{dopx}).
However, this problem is overcome
by the fact that $\delta_{IC}$ is a weak function of $\nu_{2}$,
which is contained within a logarithmic factor. Here, we have
assumed { $\nu_{2}=\rm{{ 10^{6}}}$\,GHz
(Biermann \& Strittmatter~\cite{Bie87})}.
Implicit to equation (\ref{dopx}) is the assumption that the
electrons radiating the synchrotron emission have a power-law
energy distribution ($N(E)=N_{0}E^{2\alpha-1}$, with
$h \nu_{m}\aplt E \aplt h \nu_{2}$) and that the
{ magnetic field is tangled in the rest frame of the
emitting plasma}.

{ By taking into account the dependence of $\theta_{v}$
on $\delta$ in (\ref{dopx}), $\delta_{\rm{IC}}$ can be written
as:}

   \begin{equation}
   \label{dopx2}
   \delta_{\rm{IC}} \apgt \left\{ \frac
               {{{\rm{ln}}({\nu_{2}/\nu_{m}})
	       \nu_{\gamma}}^{\alpha}
	       \left[  f ( \alpha ) S_{m} (1+z) \right]^{ ( 4 - 2 \alpha ) }}
               { S_{\gamma}{ \Theta_{v} }^{ ( 6 - 4 \alpha )}
	       \nu_{m} ^{ ( 5 - 3 \alpha ) } }
               \right\}^{1/(10 - 6 \alpha)} \,,
   \end{equation}
{ where we have defined $\Theta_{v}\equiv 3.56
\times 10^{-4} t_{v} {d_{\rm L}}^{-1} (1+z)\,\rm{mas}$.}

The results of the simultaneous INTEGRAL soft $\gamma$-ray observations
of \object{0716+714} reported by Ostorero et al. (\cite{Ost06})
{($F_{3-35\,\rm{keV}}<6.12\times 10^{-12}$\,erg\,cm$^{-2}$\,s$^{-1}$,
$F_{15-40\,\rm{keV}}<1.41 \times 10^{-11}$\,erg\,cm$^{-2}$\,s$^{-1}$,
$F_{40-100\,\rm{keV}}<1.72 \times 10^{-11}$\,erg\,cm$^{-2}$\,s$^{-1}$ and
$F_{100-200\,\rm{keV}}<5.73 \times 10^{-11}$\,erg\,cm$^{-2}$\,s$^{-1}$;}
{ see also Table~\ref{physpar} for the corresponding flux densities in
Jy), together with expression (\ref{dopx2}),} allow us to set new
constraints on the Doppler factor of \object{0716+714}
during our observations.
Assuming { $t_{v}=\rm{{ 11.1}}$\,days as in \S~\ref{Tb}},
we derive lower limits of
$\delta_{\rm{IC},8\,\rm{keV}}>\rm{{ 15.8}}$,
$\delta_{\rm{IC},23\,\rm{keV}}>\rm{{ 14.6}}$,
$\delta_{\rm{IC},63\,\rm{keV}}>\rm{{ 15.1}}$ and
$\delta_{\rm{IC},141\,\rm{keV}}>\rm{{ 14.0}}$ from the
upper limits of the flux densities of the source at 8\,keV,
23\,keV, 63\,keV and 141\,keV, respectively.

It is worth to note that, { for $\alpha=-0.3$, expression
(\ref{dopx2}) is a weak function of the redshift and the luminosity
distance of the source ($\delta_{\rm{IC}}\propto (1+z)^{-0.2}
{{d_{\rm L}}}^{0.6}$). Hence, the constraint on the Doppler
factor of the source provided by (\ref{dopx2}),
$\delta_{\rm{IC}}>\rm{{ 14.0}}$, is more robust than those
derived from expression (\ref{TB}) and the limits for $T_{\rm{B}}$
(\S~\ref{Tb}).}

\begin{table}
\caption[]{Physical parameter estimates. See the text for definitions.}
\begin{flushleft}
\small
\begin{tabular} {lcc}
\hline\noalign{\smallskip}
Parameter &  Value & Assumptions \\
\hline\noalign{\smallskip}
\multicolumn{3}{c}{ For $z=0.3$,  $d_{\rm L}=1510$\,Mpc, { $S_{m}=\rm{{ 5.23}}$\,Jy, $t_{v}=\rm{{ 11.1}}$\,days}}\\
\hline\noalign{\smallskip}
$T_{\rm{B}}^{\rm{app}}$   &    $>1.4 \times {10}^{14}$\,K    &                        $\delta = 1$\\
$\delta_{T_{\rm{B}}}$     &                   $>5.2$          & $T_{\rm{B,IC}}^{\rm{lim}}=10^{12}$\,K\\
$\delta_{T_{\rm{B}}}$     &                   $>7.8$          & $T_{\rm{B,IC}}^{\rm{lim}}=3\times 10^{11}$\,K\\
$\theta_{T_{\rm{B}}}$     &                $ <0.03$\,mas       & $\delta>7.8$         \\
$\theta_{VLBI}$      &                $ <0.04$\,mas       & 86\,GHz VLBI beam-size         \\
$\theta_{v}$         &               $ <0.03$\,mas        &                \\
$\delta_{\rm{IC},8\,\rm{keV}}$   &                   $>\rm{{ 15.8}}$          & $S_{8\,\rm{keV}}<  \rm{{  1.37}} \times 10^{-7}$\,Jy, $\alpha=-0.3$  \\
$\delta_{\rm{IC},23\,\rm{keV}}$  &                   $>\rm{{ 14.6}}$          & $S_{23\,\rm{keV}}< \rm{{  2.55}} \times 10^{-7}$\,Jy, $\alpha=-0.3$  \\
$\delta_{\rm{IC},63\,\rm{keV}}$  &                   $>\rm{{ 15.1}}$          & $S_{63\,\rm{keV}}< \rm{{  1.23}} \times 10^{-7}$\,Jy, $\alpha=-0.3$  \\
$\delta_{\rm{IC},141\,\rm{keV}}$ &                   $>\rm{{ 14.0}}$          & $S_{141\,\rm{keV}}<\rm{{  2.42}} \times 10^{-7}$\,Jy, $\alpha=-0.3$  \\
\noalign{\smallskip}
\hline
\end{tabular}
\end{flushleft}
\label{physpar}
\end{table}

\section{Conclusions}

We have presented the results from millimetre observations
of \object{0716+714}, which were performed on 2003 November 10 to 18 with
the IRAM 30\,m telescope.
Our observation strategy, based on the rapid time sampling of both
the target source and the calibrators, enabled us to reach a
relative calibration { accuracy} of $1.2$\,\% at 86\,GHz, which
demonstrates the good performance of this telescope and its ability
for future accurate IDV studies in the millimetre range.

{ During our first four observing days, the source displayed
large amplitude ($Y=34$\,\%) and monotonous inter-day variability
at 86\,GHz.
As such large amplitude variability is not expected to be
produced by standard ISS at mm wavelengths (Rickett et al.
\cite{Ric95}), this variation should be considered as
intrinsic to the source and not due to the influence of
the interstellar medium.
The similar 32\,GHz and 37\,GHz behaviour reported by
Ostorero et al.~(\cite{Ost06}) during the same observing
time range could hence be explained by intrinsic causes
also.

Ostorero et al.~(\cite{Ost06}) also report clear evidence
of IDV in the optical range, which is not matched either
in the mm or in the cm bands (see also Fuhrmann et al.~\cite{Fuh06}).
This indicates that the radio-mm emitting region was
located in a different region than the optical one or
that their radiation behaviours were driven by different
physical processes during our observations.}

We have reported { an unusually large linear
polarization degree $<p>=\rm{(}15.0\pm1.8\rm{)}\,\%$
of \object{0716+714} at 86\,GHz, which suggests a large level
of magnetic field alignment.
{ At such frequency, linear polarization inter-day
variability, with significance level $\apgt 95\,\%$;
$\sigma_{P}/<P>=15$\,\% and $\sigma_{\chi}=6^{\circ}$,
was observed during the first four observing days.}
We have also shown} $2 \sigma$ evidence of { simultaneous}
polarization flux density and polarization angle IDV within
a time $\aplt 24$\,h.
If such rapid polarization variations that are uncorrelated
with the total flux density variability are confirmed by future
observations, then equally rapid changes of the magnetic field
configuration of the source or changes of opacity around $\nu_{m}$
would be required to explain the phenomenon.
In both cases, inhomogeneous models would probably
have to be invoked.

The synchrotron spectrum of \object{0716+714} peaked at
$\nu_{m} \approx 86$\,GHz during 2003 November 10 to 15
{ (Ostorero et al.~\cite{Ost06})}, and { showed}
an optically thin spectral index $\bar{\alpha}_{86-229}=-0.23 \pm 0.10$
between 86\,GHz and 229\,GHz. The apparent brightness temperature
derived from our 86\,GHz light curve, $T_{\rm{B}}^{\rm{app}}>1.4 \times
{10}^{14}$\,K for a redshift $z=0.3$, exceeds, at least by
two orders of magnitude, the IC-limit of $3\times {10}^{11}$\,K
(Kellemann \& Pauliny-Toth~\cite{Kel69}; Readhead~\cite{Rea94};
Kellermann~\cite{Kel03}).
This mismatch can be explained by the relativistic motion or
expansion of the source with a minimum Doppler factor
$\delta > 7.8$.
The upper limits from the soft $\gamma$-ray simultaneous
INTEGRAL observations in the 3\,keV to 200\,keV energy range
enabled us to compute { an } independent and more { robust} limit
for the source Doppler factor, $\delta_{\rm{IC}} \apgt \rm{{ 14}}$.
This limit is consistent with previous estimates { of}
the Doppler factor of \object{0716+714} measured from the kinematics
of VLBI-scale jet features (Bach et al.~\cite{Bac05}),
which ranged from $\delta_{\rm{VLBI}}=20$ to $\delta_{\rm{VLBI}}=30$.
Such high Doppler factors have also been measured
with 43\,GHz-VLBI by Jorstad et al.~(\cite{Jor05}), who
detected $\delta_{\rm{VLBI}}>20$ in 9 of the 13 monitored blazars.

{ As} no soft-$\gamma$-ray IC-avalanches were detected
by INTEGRAL during our observations and the reported large
amplitude 86\,GHz variability can not be ascribed to the ISS,
the relativistic beaming of the radiation coming from the
emitting region in \object{0716+714} { offers} a robust explanation
of the { apparent} violation of the IC-limit
{ of the brightness temperature in the mm range}.
Note, however, that (total or partial) coherent
synchrotron-emission scenarios can not be ruled out.

Finally, we should stress that we have proven that \object{0716+714},
in particular, and blazars, in general, can display
apparent brightness temperatures two orders of magnitude larger
than the theoretical limits { in the millimetre range} and
that the influence of the interstellar medium is not always necessary
to explain the mismatch between observations and theory.
Following the above arguments, we are rather confident that, apart
from inaccuracies in our assumptions, the estimates and limits
for the physical parameters of \object{0716+714} reported in the previous
sections correspond to those governing the observed source
behaviour.


\begin{acknowledgements}
      We gratefully acknowledge A.~L. Roy, A.~P. Marscher
      and A.~P. Lobanov for their helpful suggestions on this
      paper, { C. Thum and H. Wiesemeyer
      for their useful comments and information on polarization
      calibration of IRAM 30\,m telescope data and} A. Sievers for
      his help in the calibration of { our} data.
      { We also wish to acknowledge helpful comments by the
      anonymous referee.}
      I. Agudo, E. Angelakis, L. Fuhrmann, U. Bach, L. Ostorero
      and J. Gracia acknowledge financial support from the EU
      Commission { under contract} HPRN-CT-2002-00321
      (ENIGMA network).
      This paper is based on observations carried out at the IRAM
      30\,m telescope. IRAM is supported by INSU/CNRS (France),
      MPG (Germany) and IGN (Spain).
\end{acknowledgements}


\appendix

\section{Response of the IRAM 30\,m heterodyne
         receivers to a partially linearly polarized wave}
\label{appA}

The power received by an antenna when it observes a
source of arbitrary polarization can be characterised as

   \begin{equation}
   \label{krausform}
   W=\frac{1}{2} S_{0} A_{e}(1-d) +
   S_{0} A_{e} d \cos^{2} \frac{\gamma} {2} \,,
   \end{equation}
where $S_{0}$ is the total flux density of the source,
$A_{e}$ is the effective aperture of the antenna,
$d$ the source degree of polarization and $\gamma$ the
angle between the wave and the antenna polarization states
on the Poincar\'e sphere (see e.g. Kraus~\cite{Kra86}).
The first and second terms on the right-hand side of Eq.
(\ref{krausform}) represent the unpolarized and polarized
responses of the antenna, respectively.

\begin{figure}
   \centering
   \includegraphics[bb=0 0 368 426,width=7cm,clip]{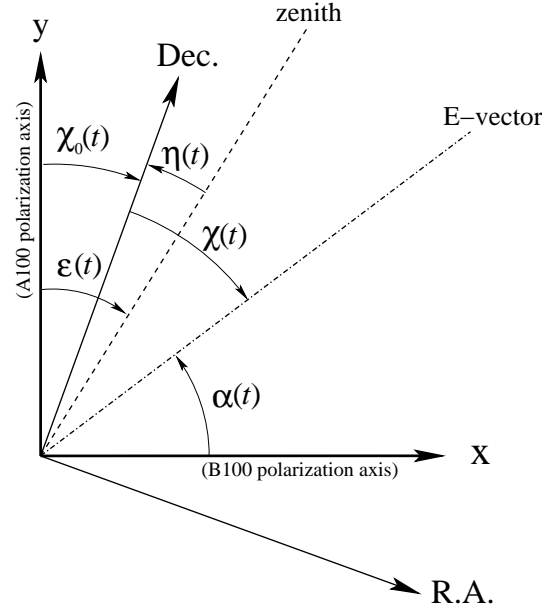}
      \caption{Angles in the reference polarization plane ({\it x},
      {\it y}) as seen from the A100 and B100, IRAM 30\,m
      heterodyne receivers. $\alpha(t)$, which is defined in
      ({\it x}, {\it y}) and can be measured by fitting
      equations (\ref{polA100minus}) and (\ref{polB100minus}), relates
      to the astronomical { electric vector position}
      angle $\chi(t)$ (defined from north
      to east in the equatorial coordinate system (R.A.,Dec.)), to the
      parallactic angle $\eta(t)$ and to the elevation angle $\epsilon(t)$.}
   \label{chisys}
\end{figure}

If the polarized component of the incident radiation is
only linearly polarized, then $d$ is the degree of
linear polarization ($p$).
In addition, for the case of the IRAM 30\,m millimetre
radio telescope, when heterodyne receivers are used,
$\gamma \equiv \alpha+\beta$.
Here, $\alpha$ is the angle between the polarization
direction of the incident signal and the horizontal
direction in the telescope receiver cabin
-the $x$ axis of the reference polarization plane,
see Fig. \ref{chisys}-.
$\beta$ is the angle between the $x$ axis and the
polarization orientation of the receiver.
For the particular case of receivers A100 and B100,
$\beta$ is $\pi$/2 and 0, respectively.

Taking into account that $W=k T_{A}^{*}$ and
$T_{A}^{*}/S_{obs}=A_{e}/(2 k)$, where $k$ is the
Boltzmann constant, $T_{A}^{*}$ is the calibrated
antenna temperature outlined on \S \ref{sred} and $S_{obs}$
is an equivalent total flux density observed for the receiver,
Eq. (\ref{krausform}) may be expressed for A100 and B100 as:

   \begin{equation}
   \label{A100form}
   S_{\rm{A100}}(t)=
   S_{0}(t) (1-p(t)) +
   2 S_{0}(t) p(t) \sin^{2} (\alpha(t)) \,
   \end{equation}
and
   \begin{equation}
   \label{B100form}
   S_{\rm{B100}}(t)=
   S_{0}(t) (1-p(t)) +
   2 S_{0}(t) p(t) \cos^{2} (\alpha(t)) \,.
   \end{equation}
where we have introduced the possible time dependence
of each variable and we have defined $S_{\rm{A100}}(t)
\equiv S_{obs}(\alpha(t),\beta=\pi/2,t)$ and
$S_{\rm{B100}}(t) \equiv S_{obs}(\alpha(t),\beta=0,t)$.
From (\ref{A100form}) and (\ref{B100form}) and defining
the linearly polarized flux density of the observed source
as $P \equiv S_{0} p$, then it is straightforward to show
that:

   \begin{equation}
   \label{S_medform}
   \frac {S_{\rm{A100}}(t) + S_{\rm{B100}}(t)}{2}
   =S_{0}(t) \,,
   \end{equation}

   \begin{equation}
   \label{polA100minus}
   S_{\rm{A100}}(t)
   - S_{0}(t)
   = P(t) \cos{2\left(\alpha(t)+\frac{\pi}{2}\right)} \,
   \end{equation}
and
   \begin{equation}
   \label{polB100minus}
   S_{\rm{B100}}(t)
   - S_{0}(t)
   =P(t) \cos{2(\alpha(t))} \,,
   \end{equation}

Thum et al.~\cite{Thu00} showed that, for the IRAM 30\,m telescope,
$\alpha(t)$ can be expressed as the following function of
astronomical angles:

   \begin{equation}
   \label{angdef}
   \alpha(t)=\frac{\pi}{2}-\chi(t)+\eta(t)-\epsilon(t) \,,
   \end{equation}
(see also Fig.~\ref{chisys}) where $\chi$(t) is the { electric vector
position} angle of the source (defined from north to east in the equatorial
coordinate system (R.A.,Dec.)) and $\eta(t)$ and $\epsilon(t)$ are,
respectively, the parallactic angle and elevation angles
(see Fig. \ref{chisys} and, Thum et al.~\cite{Thu00}).
Note that the location of the receiver cabin in the
Nasmyth focus of the IRAM 30\,m telescope causes a
rotation of the equatorial reference system by an angle
$\chi_{0}(t)=\epsilon(t)-\eta(t)$ with regard to the
polarization reference system ({\it x}, {\it y}).

\section{Statistical formulation for variability}
\label{appB}

The statistical variables used in this paper are
defined in detail by Quirrenbach et al. (\cite{Qui00})
and Kraus et al. (\cite{Kra03}). Here we present a summary
of { these} definitions.

The modulation index,

   \begin{equation}
      \label{m}
      m[\%]=100 \frac{\sigma_{S}}{<S>} \,,
   \end{equation}
(where $\sigma_{S}$ is the standard deviation of the flux
density measurements and ${<S>}$ is its average in time)
provides the strength of the observed { variability} amplitude
of each source without taking into account the uncertainty
of the measurements.

The {\it variability amplitude},

   \begin{equation}
      \label{Y}
      Y[\%]=3 \sqrt{m^{2}-m_{0}^{2}} \,,
   \end{equation}
(where $m_{0}$ is the modulation index of the non-variable
and nearby calibrators) gives a measurement of the variability
amplitude with regard to that of the non-variable calibrators,
for which $Y$ is set to zero.

The reduced chi-squared,

   \begin{equation}
      \label{Y}
      \chi_{r}^2=\frac{1}{N-1} \sum_{i=1}^N
                 \left(\frac{S_i-<S>}{\Delta S_{i}}\right)^{2} \,,
   \end{equation}
(where $N$ is the number of measurements, the $S_i$ are the
individual flux densities and $\Delta_i$ their errors) tests
the hypothesis that the measured light curves could be modelled
by a constant function. We consider that a source is variable
(with a significance level $\geq 99.9$\,\%) only if its light
curve shows a probability to be constant $\leq 0.1$\,\%.

Finally, the structure function is defined as

   \begin{equation}
      \label{Y}
      D(\tau)={<(S(t)-S(t+\tau))^{2}>} \,,
   \end{equation}
(with $< >$ denoting averaging in time).
A characteristic time-scale in the light curves (i.e., the time
between a maximum and a minimum or vice-versa) causes a maximum in
$D(\tau)$, while a periodic pattern produces a minimum.


\end{document}